\documentclass[]{pasj01}

\Received{}
\Accepted{}
 
\let\uproot\relax
\let\leftroot\relax

\usepackage{amsmath}
\usepackage{hyphenat} 
\usepackage{url}
\usepackage{bm}
\usepackage{multirow}
\usepackage{lineno}

\begin{document}

\title{ 
Where's Swimmy?: Mining unique color features buried in galaxies by deep anomaly detection using Subaru Hyper Suprime-Cam data}

\author{Takumi S. \textsc{Tanaka}\altaffilmark{1,2}$^\ast$}
\altaffiltext{1}{Department of Astronomy, School of Science, The University of Tokyo, 7-3-1 Hongo, Bunkyo-ku, Tokyo 113-0033, Japan}
\email{tanaka-takumi949@g.ecc.u-tokyo.ac.jp (TT)}

\author{Rhythm \textsc{Shimakawa}\altaffilmark{2}$^\ast$}
\altaffiltext{2}{National Astronomical Observatory of Japan (NAOJ), National Institutes of Natural Sciences, Osawa, Mitaka, Tokyo 181-8588, Japan}
\email{rhythm.shimakawa@nao.ac.jp (RS)}

\author{Kazuhiro \textsc{Shimasaku}\altaffilmark{1,3}$^\ast$}
\altaffiltext{3}{Research Center for the Early Universe, The University of Tokyo, 7-3-1 Hongo, Bunkyo-ku, Tokyo 113-0033, Japan}
\email{shimasaku@astron.s.u-tokyo.ac.jp (KS)}

\author{Yoshiki \textsc{Toba}\altaffilmark{4,5,6}}
\altaffiltext{4}{Department of Astronomy, Kyoto University, Kitashirakawa-Oiwake-cho, Sakyo-ku, Kyoto 606-8502, Japan}
\altaffiltext{5}{Academia Sinica Institute of Astronomy and Astrophysics, 11F of Astronomy-Mathematics Building, AS/NTU, No.1, Section 4, Roosevelt Road, Taipei 10617, Taiwan}
\altaffiltext{6}{Research Center for Space and Cosmic Evolution, Ehime University, 2-5 Bunkyo-cho, Matsuyama, Ehime 790-8577, Japan}

\author{Nobunari \textsc{Kashikawa}\altaffilmark{1,3}}

\author{Masayuki \textsc{Tanaka}\altaffilmark{2}}

\author{Akio K. \textsc{Inoue}\altaffilmark{7,8}}
\altaffiltext{7}{Department of Physics, School of Advanced Science and Engineering, Faculty of Science and Engineering, Waseda University, 3-4-1 Okubo, Shinjuku, Tokyo 169-8555, Japan}
\altaffiltext{8}{Waseda Research Institute for Science and Engineering, Faculty of Science and Engineering, Waseda University, 3-4-1 Okubo, Shinjuku, Tokyo 169-8555, Japan}


\KeyWords{galaxies: general --- galaxies: nuclei --- galaxies: statistics}

\maketitle

\begin{abstract}
We present the {\sc Swimmy} (Subaru WIde-field Machine-learning anoMalY) survey program, a deep-learning-based search for unique sources using multicolored ($grizy$) imaging data from the Hyper Suprime-Cam Subaru Strategic Program (HSC-SSP). 
This program aims to detect unexpected, novel, and rare populations and phenomena, by utilizing the deep imaging data acquired from the wide-field coverage of the HSC-SSP. 
This article, as the first paper in the {\sc Swimmy} series, describes an anomaly detection technique to select unique populations as “outliers" from the data-set.
The model was tested with known extreme emission-line galaxies (XELGs) and quasars, which consequently confirmed that the proposed method successfully selected ~60$\sim$70\% of the quasars and 60\% of the XELGs without labeled training data. 
In reference to the spectral information of local galaxies at $z=$ 0.05--0.2 obtained from the Sloan Digital Sky Survey, we investigated the physical properties of the selected anomalies and compared them based on the significance of their outlier values.
The results revealed that XELGs constitute notable fractions of the most anomalous galaxies, and certain galaxies manifest unique morphological features.
In summary, deep anomaly detection is an effective tool that can search rare objects, and ultimately, unknown unknowns with large data-sets.
Further development of the proposed model and selection process can promote the practical applications required to achieve specific scientific goals.
\end{abstract}

\section{Introduction}

Multiple intensive surveys have successfully unveiled the broad outline of 
galaxy formation and evolution; however, certain essential factors still remain unidentified 
(e.g., \cite{Bland2016,Kormendy2013,Naab2017,Wechsler2018}).
In addition, we cannot expunge the possibility that we are still missing certain populations and events, referred to as “unknown unknowns". 

Recent studies demonstrate that rare objects such as extreme emission line galaxies (XELGs), low luminosity quasi-stellar objects (quasar), and dual quasars could be the 
missing pieces in the complete representation of galaxy formation and evolution.
In context, XELGs are presumed to be analogs of high-$z$ galaxies that provide information regarding the drivers of cosmic reionization 
\citep{Cardamone2009,Amorin2015,Greis2016,Yang2017}.
Additionally, low-luminosity quasars aid in characterizing the relationship between host galaxies and their quasars 
\citep{Salim2007,Schawinski2010,Matsuoka2014,Trump2015,Ishino2020,Li2021},
and dual quasars are supposedly in a late-stage merger and provide information regarding the relationship between mergers and supermassive black hole (SMBH) evolution
\citep{Hopkins2008,Blecha2018,Silverman2020,Tang2021}. 
Further discoveries of rare populations and unknown unknowns would aid in the detailed understanding of galaxy formation and evolution. 
However, searching for such uncommon sources requires a massive and expensive data-set. 
Notwithstanding the large data, the efficient collection of such sources is a challenge that prevents us from statistically investigating their detailed characteristics. 
Therefore, developing a method of an exhaustive search for rare objects and unknown unknowns based on extensive data-sets is quite important.

From the data-set perspective, ongoing and upcoming wide-field deep surveys using large aperture telescopes are currently exploring rare populations on an advanced level. 
The Hyper Suprime-Cam Subaru Strategic Program (HSC-SSP: \cite{Aihara2018}) is a large survey project, which uses the Hyper Suprime-Cam (HSC: \cite{Miyazaki2018}) operating on the Subaru telescope.
The HSC-SSP comprises three layers: wide, deep, and ultradeep; the wide layer covers 1400 deg$^2$ with $grizy$ multicolored images.
In comparison to the Sloan Digital Sky Survey (SDSS, \cite{York2000}), the HSC-SSP provides deeper images owing to its larger 8.2-m-aperture mirror and longer exposure periods \citep{Furusawa2018,Kawanomoto2018,Komiyama2018}.
Such high-quality imaging data enables the detection of faint rare objects (e.g., extremely metal-poor galaxies (XMPGs): \cite{Kojima2020}, high-redshift low-luminosity quasars: \cite{Matsuoka2019}, gravitationally-lensed objects: \cite{Sonnenfeld2018}, dust-obscured galaxies: \cite{Toba2015} ,and high-$z$ protoclusters: \cite{Harikane2019}).

From the perspective of methodology, anomaly detection has recently garnered attention in detecting extremely rare sources from large data-sets. 
Anomaly detection is a technique that is used to identify uncommon features in sources in a given data-set, which are significantly distinct from most of the existing sources, without using any labeled training data; it means the anomaly detection method can detect unknown anomalous objects or more anomalous objects than known objects or assumed templates, which may be missed by supervised classifiers reliant on certain labeled training data or templates. 
Thus, the anomaly detection does not require large data-sets of the target rare populations in advance.
This novel approach has impacted a wide range of fields such as pathogen detection in magnetic resonance images \citep{Baur2021}, diagnoses of amplitude anomalies from seismic waveform data \citep{Zaccarelli2021}, and a search for new particles in particle physics \citep{Nachman2020}. 
This method is also useful for searching rare objects and unknown unknowns in astronomical sources in a large data-set. 
Indeed, based on the big library from the SDSS, various statistical attempts have been made for astronomical anomaly detection, such as
Bayesian algorithms \citep{Mortlock2012}, self-organizing maps \citep{Meusinger2012,Fustes2013}, clustering algorithms \citep{Andrae2010,Sanchez2013}, random forest algorithms \citep{Baron2017,Hocking2018}, support vector machines \citep{Solarz2017}, and deep generative models \citep{Stark2018,Margalef2020}. 
A recent study by \citet{Storey-Fisher2021} applied deep generative model to imaging data from the second data release of the HSC-SSP. 
However, a majority of the studies elaborated in prior research were experimental, and thus, greater practical scientific applications are desired with anomaly detection methods.

With this motivation, we launched the {\sc Swimmy} (Subaru WIde-field Machine-learning anoMalY) survey program, which aims at a blind search of extremely unique galaxy populations in large data-sets obtained from the HSC-SSP and the Subaru Strategic Programs with Prime Focus Spectrograph (PFS-SSP, \cite{Takada2014,Tamura2016}). 
This paper, as the first of this series, reports the development of a simple anomaly detection model based on machine learning, which was applied on the HSC-SSP multicolor images to detect outlier objects that exhibit anomalous features primarily in the central region of galaxies.
The proposed model was tested by employing known quasar and XELG samples. 
As they respectively host active SMBHs in the centers and emission-line dominant regions, the anomaly detection method can identify such uncommon features as outliers. 
Unlike prior anomaly detection surveys, this study scientifically examined the physical characteristics of the selected anomaly sources by combining archive data to provide useful information for further development of anomaly detection methods. 
Such an effort is required for the forthcoming enormous data that would be delivered by the next-generation legacy surveys conducted in the Vera C. Rubin Observatory. 
The Rubin Observatory will provide us with the largest imaging data-set covering 18~000 $\mathrm{deg}^2$ and 20 billion galaxies \citep{Ivezic2019}, which is adequately large to detect a $6.5\sigma$ anomaly by simple arithmetic, approximately one out of twelve billion.

The remainder of this paper is organized as follows.
The data, sample selection, and data construction are described in section~\ref{s2}, after which the machine-learning based anomaly detection model is presented in section~\ref{s3} along with the selection of the anomaly candidates from the data-set.
In section~\ref{s4}, we discuss the physical properties of the detected anomaly objects from spectral data, and lastly the results are summarized in section~\ref{s5}. 
Throughout this paper, the AB magnitude system \citep{Oke1983} was adopted, assuming a \citet{Kroupa2001} initial mass function and a flat WMAP7 cosmology, $H_0=70$ km~s$^{-1}$ Mpc$^{-1}$, $\Omega_m=0.27$ \citep{Komatsu2011}.

\section{Data}\label{s2}

\subsection{HSC-SSP data}\label{s21}

This study employs multi-band ($grizy$) imaging data from the S20A internal release of the HSC-SSP (wide layer), which is presently available as the third public data release (HSC-SSP PDR3; \cite{Aihara2021}). 
All the images were reduced using a dedicated pipeline, {\tt hscPipe} \citep{Bosch2018}. 
The HSC-SSP survey field is divided into $\sim1.7\times1.7$ square degree areas called {\tt tracts} and further subdivided into $\sim12\times12$ square arcmin areas called {\tt patches}. 
We excluded the patches with full-width-half-maximum (FWHM) larger than 1.0 arcsec or shallow imaging depth of the bottom 5th percentiles in PSF limiting magnitude in one or more filters.
The thresholds of the $5\sigma$ PSF limiting magnitude were $g<25.82$, $r<25.49$, $i<25.23$, $z<24.47$, or $y<23.59$. 
After removing the inferior-quality patches, we obtained images of 815 square degrees.

\subsection{Training and test data-set}\label{s22}

\begin{figure}
 \begin{center}
  \includegraphics[width=7.5cm]{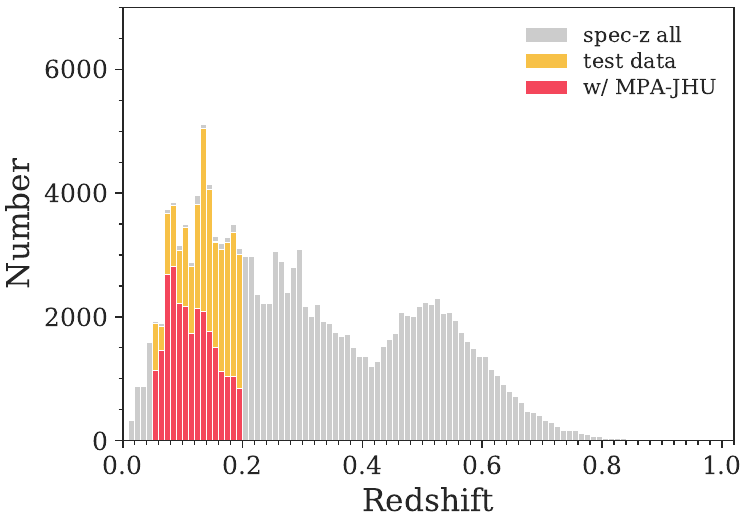} 
 \end{center}
\caption{
Redshift distribution of spec-$z$ sources ($i<20$ mag) available in the HSC-SSP PDR3 library (gray), galaxies in the test data used in this study (yellow), and galaxies cross-matched with the MPA-JHU catalog (red; refer text for details). 
}
\label{specz_dist}
\end{figure}

We restrict our analysis to 49~319 luminous galaxies in the above area that were brighter than 20 mag in the $i$-band {\tt cmodel} magnitude (composite model; \cite{Abazajian2004,Bosch2018}) and within a redshift range of $z=0.05$--0.2 (figure~\ref{specz_dist}), where spectroscopic redshifts are obtained by looking for counterparts in the SDSS DR15 \citep{Blanton2017,Aguado2019,Bolton2012} and the GAMA DR3 \citep{Baldry2018,Hopkins2013,Baldry2014} with a search radius of 1 arcsec \citep{Aihara2021}. 
In addition, we removed sources that were prominently affected by nearby stars, cosmic rays, corrupted pixels, or saturated pixels by referring to various flags ({\tt is False}) stored in the database:
{\tt pixelflags\_edge}, {\tt pixelflags\_interpolatedcenter}, 
{\tt pixelflags\_saturatedcenter}, {\tt pixelflags\_crcenter}, 
{\tt pixelflags\_bad}, {\tt sdssshape\_flag}, 
{\tt pixelflags\_bright\_objectcenter}, {\tt mask\_brightstar\_halo}, 
{\tt mask\_brightstar\_ghost}, and {\tt mask\_brightstar\_blooming} 
\citep{Coupon2018,Aihara2021,Bosch2018}. 
Moreover, the point sources were excluded by imposing {\tt psfflux\_mag$-$cmodel\_mag$>$0.2} in the $i$ band (see also \cite{Strauss2002,Baldry2010}).

Subsequently, we categorized the sample into several classes to train and test our selection model. 
First, we cross-matched the sample with the MPA-JHU catalog \citep{Brinchmann2004,Kauffmann2003,Tremonti2004}, which summarizes detailed spectral measurements for the SDSS DR7 spectra \citep{Abazajian2009} and is useful for constructing training data and investigating the physical properties of anomaly candidates. 
In the anomaly selection, it is essential to exclude uncommon sources from the training data as many as possible to ensure that the proposed model does not learn outlier features from the training sample. 
Thus, we established a data-set based on the MPA-JHU catalog, excluding known XELGs, quasars, and AGN candidates. 
Herein, the XELG galaxies were defined with the observed equivalent width (EW) of either [O{\sc ii}]$\lambda\lambda$3726,3729, H$\beta$+[O{\sc iii}]$\lambda\lambda$4959,5007, or H$\alpha$+[N{\sc ii}]$\lambda\lambda$6548,6584, being larger than 100 \AA, and 701 sources satisfied this definition. 
In addition, we detected 23 sources that have a counterpart in the latest quasar catalog (DR16Q v4: \cite{Lyke2020}). 
The XELG and quasar samples were used to test the performance of the proposed anomaly selection model, as discussed in subsection~\ref{s41}. 
In addition, possible AGN and low-ionization nuclear emission-line regions (LINER) candidates were removed from the training data based on the flag ({\tt BPTCLASS} $\ge3$) in the SDSS library, which classifies galaxy spectra using the Baldwin, Phillips, \& Terlevich (BPT) diagram \citep{Baldwin1981,Kauffmann2003}.
Based on these selection criteria, we established the training data that amounted to $N=$ 21~151 (table~\ref{data_summary}). 
Again the training data should ideally not contain anomalies; however, the training data may have undetected anomaly objects or unknown unknowns because these selection criteria depended on only the SDSS 1-D spectral data.
Additionally, although this study does not delve into the remaining 23~582 spec-$z$ sources, they were still adopted for use in anomaly detection as well as the MPA-JHU sample. 

Moreover, we constructed color-selected samples without spectroscopic counterparts for a complementary analysis, as detailed in appendix~\ref{A3}.

\begin{table}
  \caption{
  {Summary of the data used in this study.}
  }{%
  \begin{tabular}{lr}
      \hline
      Data & N \\ 
      \hline
      $i<20$ galaxies at $z=0.05$--0.2 (test data)  & 49~319\\
      \quad(1) SDSS MPA-JHU & 25~714\\
      \quad\quad(1-1) Training data & 21~151\\
      \quad\quad(1-2) XELGs & 701\\
      \quad(3) DR16Q quasars & 23\\
      \quad(4) Other spec-$z$ (SDSS \& GAMA) & 23~582 \\
      \hline
    \end{tabular}}
    \label{data_summary}
\end{table}

\subsection{Preparation of data cubes}\label{s23}

For each galaxy, we retrieved an image of $64\times64$ pixels ($11\times11$ arcsec$^2$, with a pixel scale of 0.168 arcsec per pixel: the default size in the HSC-SSP) for each of the five $grizy$ broad-bands.
Thereafter, we scaled the pixel values to a range of [0, 1] for the proposed deep-learning algorithm.
There are various ways of flux normalization such as logarithmic, power-law, or square-root stretches. 
Although the appropriate selection of a normalization formula for deep learning of galaxy imaging data is still debatable, this research followed the approach by \citet{Lupton1999,Lupton2004} that adopts an inverse hyperbolic sine function (arcsinh).
Moreover, the arcsinh stretch has been commonly applied to galaxy images in public engagement programs such as {\tt Galaxy Zoo}\footnote{\url{https://www.zooniverse.org/projects/zookeeper/galaxy-zoo/}} \citep{Lintott2008,Lintott2011,Willett2013} and {\tt GALAXY CRUISE}\footnote{\url{https://galaxycruise.mtk.nao.ac.jp}}. 
More importantly, this technique can avoid color saturation as the pixel values were normalized across color channels, and thus allow the model to interpret local anomalous color features within the galaxy image. 

\begin{figure}
 \begin{center}
  \includegraphics[width=7.5cm]{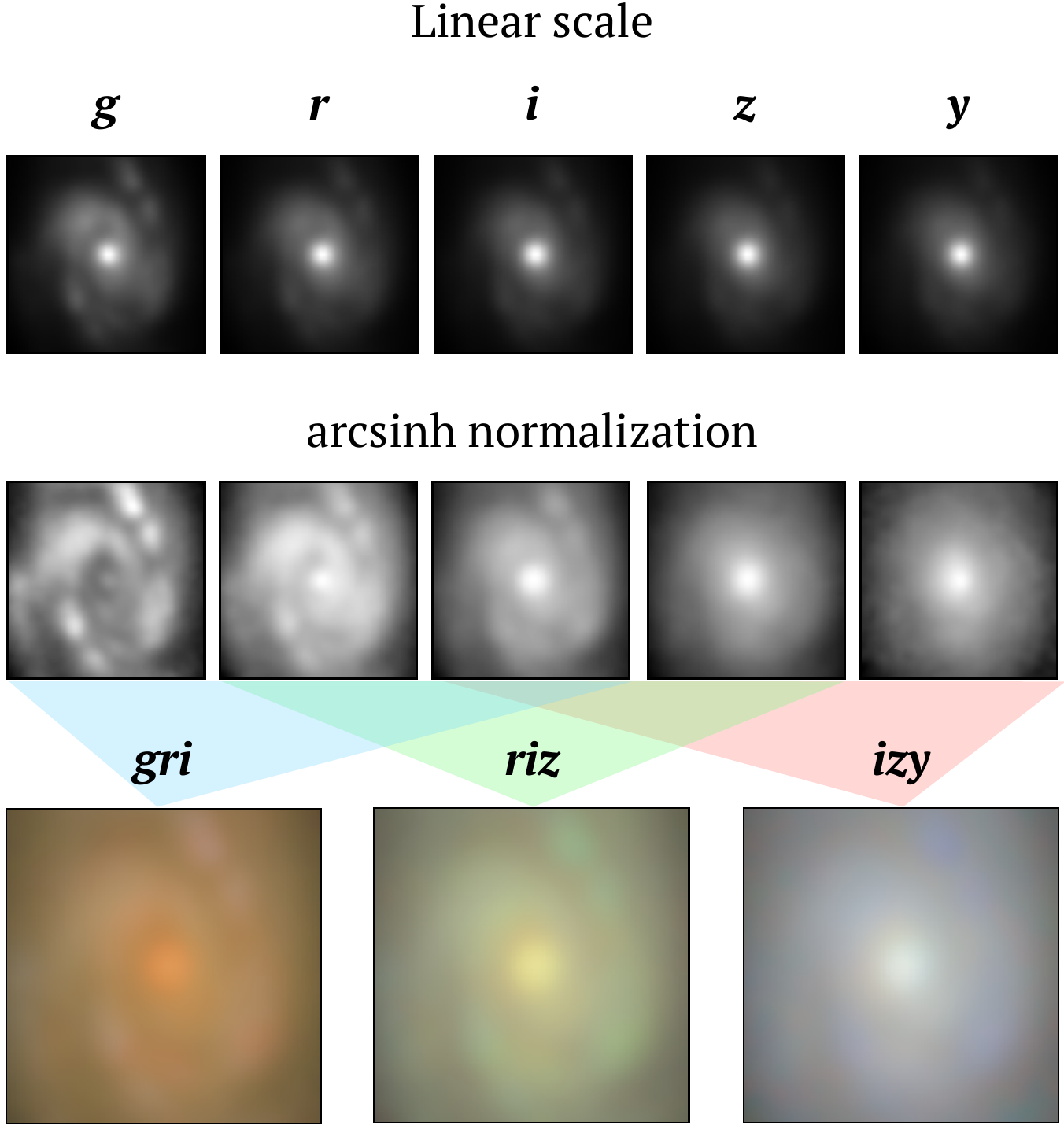} 
 \end{center}
\caption{An example of the arcsinh normalization for a spiral galaxy ({\tt object\_id=36429603367044827} at $z=0.069$). 
The top and middle panels depict the original $grizy$ images with linear scaling in individual filters and those after arcsinh normalization over five broadbands (equation~\ref{asinh}), respectively.
The bottom color images depict RGB maps based on three of the five bands (left to right: $gri$, $riz$, and $izy$). 
} 
\label{asinh_stretch}
\end{figure}

Furthermore, we slightly modified the code for arcsinh normalization written by the HSC-SSP pipeline team\footnote{\url{https://hsc-gitlab.mtk.nao.ac.jp/ssp-software/data-access-tools/tree/master/pdr3/colorPostage}} to create multicolored imagine cubes for the hscMap\footnote{\url{https://hsc-release.mtk.nao.ac.jp/hscMap-pdr3/app/}}, which is accessible on the HSC-SSP database and citizen science program, {\tt GALAXY CRUISE}.
This algorithm was compiled to apply the arcsinh stretch to five broadband images acquired from the HSC-SSP, $x\in A=\{g,r,i,z^\ast,y^\ast\}$, where $x$ is a cut-out image in each band with a zero-point magnitude of 19 mag. 
Nevertheless, we scaled the $z$ and the $y$ band data by 1/1.2 and 1/1.5, respectively, corresponding to the typical flux to the $i$ band flux ratios. 
Such additional scaling aided in improving the balance of normalized fluxes between the broadbands.
Otherwise, the proposed anomaly detection model will assign a higher weight to the $z$ and the $y$ band data, which are shallower than the $i$ band data \citep{Aihara2021}. 
The normalized flux $f(x)$ can be calculated as 
\begin{linenomath}
\begin{equation}
    f(x) = \frac{\mathrm{arcsinh}(\alpha I)}{\mathrm{arcsinh}(\alpha)\cdot xI} + \beta,
\label{eq1}
\end{equation}
\end{linenomath}
where $I\equiv{\displaystyle \Sigma_{x\in A}}/5$, and the original values were used in the hscMap for the scaling parameter, $\alpha=\exp(10)$, and the bias parameter, $\beta=0.05$.
Moreover, the flux range was limited to [0, 1], defined as follows:
\begin{linenomath}
\begin{equation}
    f(x) = \left\{ \begin{array}{ll}
    0 & (f(x)<0), \\
    f(x) & (0\leq f(x)\leq 1), \\
    1 & (1<f(x)).
  \end{array} \right.
\label{asinh}
\end{equation}
\end{linenomath}

The conducted scaling process and color visualizations based on the three bands of five broadband images are illustrated in figure~\ref{asinh_stretch}. 
The standardization using the mean flux over the entire broadband ($I$) allows us to avoid color saturation \citep{Lupton1999}.
Note that the $izy$ bands trace the stellar continua of targets at the rest-frame $\geq6000$ \AA, and the $z$ and the $y$ band fluxes are scaled to the $i$ band flux across the board, which causes an unclear color gradient in these three images.
For instance, in case there is a strong H$\alpha$ line contribution to the $i$ band flux, a significant color excess would be detected in the $i$ band (section~\ref{s4}). 
Therefore, all the images were smoothed with a Gaussian kernel prior to flux scaling to match the observed FWHM to 1.0 arcsec, based on the observed information stored in each {\tt patch} (subsection~\ref{s21}). 

\section{Methodology}\label{s3}

\subsection{Model description}\label{s31}

The denoising convolutional autoencoder (DCAE), which was applied in this study to detect anomalous galaxies, is an autoencoder (AE; \cite{Hinton2006}) equipped with a layer that add noise \citep{Vincent2008} and Convolutional Neural Networks (CNN; \cite{Lecun1998,Krizhevsky2012,Lecun2015}).
The AE is a machine-learning model that learns a latent representation of the input data $x$ using two Neural Networks (NN): an encoder and a decoder. 
The encoder $f\left(x;\bm{\theta}\right)$ with $\bm{\theta}$ as a parameter vector reduces the dimension of the input image $x$, and represents $x$ in a low-dimensional data $z=f\left(x;\bm{\theta}\right)$ in the latent space.
Subsequently, the decoder $g\left(z;\bm{\phi}\right)$with $\bm{\phi}$ as a parameter vector reconstructs the input data $x'=g\left(z;\bm{\phi}\right)$ from the encoded data $z$. 
Thereafter, the two NNs are optimized to minimize the value of the loss function $L\left(x,g\left(f\left(x\right)\right)\right)$, which is a function that evaluates the success of the learning process using the original data $x$ and reproduction data $g\left(f\left(x\right)\right)$. 
Upon considering the mean--squared error (MSE) as the loss function, the optimization relation can be expressed as
\begin{linenomath}
\begin{equation}
    \underset{\bm{\theta}, \bm{\phi}}{\mathrm{argmin}}\ \sum_i L\left(x_i,g\left(f\left(x_i\right)\right)\right) = 
    \underset{\bm{\theta}, \bm{\phi}}{\mathrm{argmin}}\ \sum_i \left(x_i-x'_i\right)^2 .
\end{equation}
\end{linenomath}

Prior to encoding, we added noise to the input data using a GaussianNoise layer that added Gaussian noise with a mean value of 0.
The standard deviation of the Gaussian noise, $r_{\rm gauss}$, is a hyperparameter, and we adopted $r_{\rm gauss}=0.02$ (i.e., Gaussian noise with a standard deviation of 2 percent to the max count) through a trial-and-error process as described in appendix~\ref{A1}. 

\begin{figure*}
 \begin{center}
  \includegraphics[width=16cm]{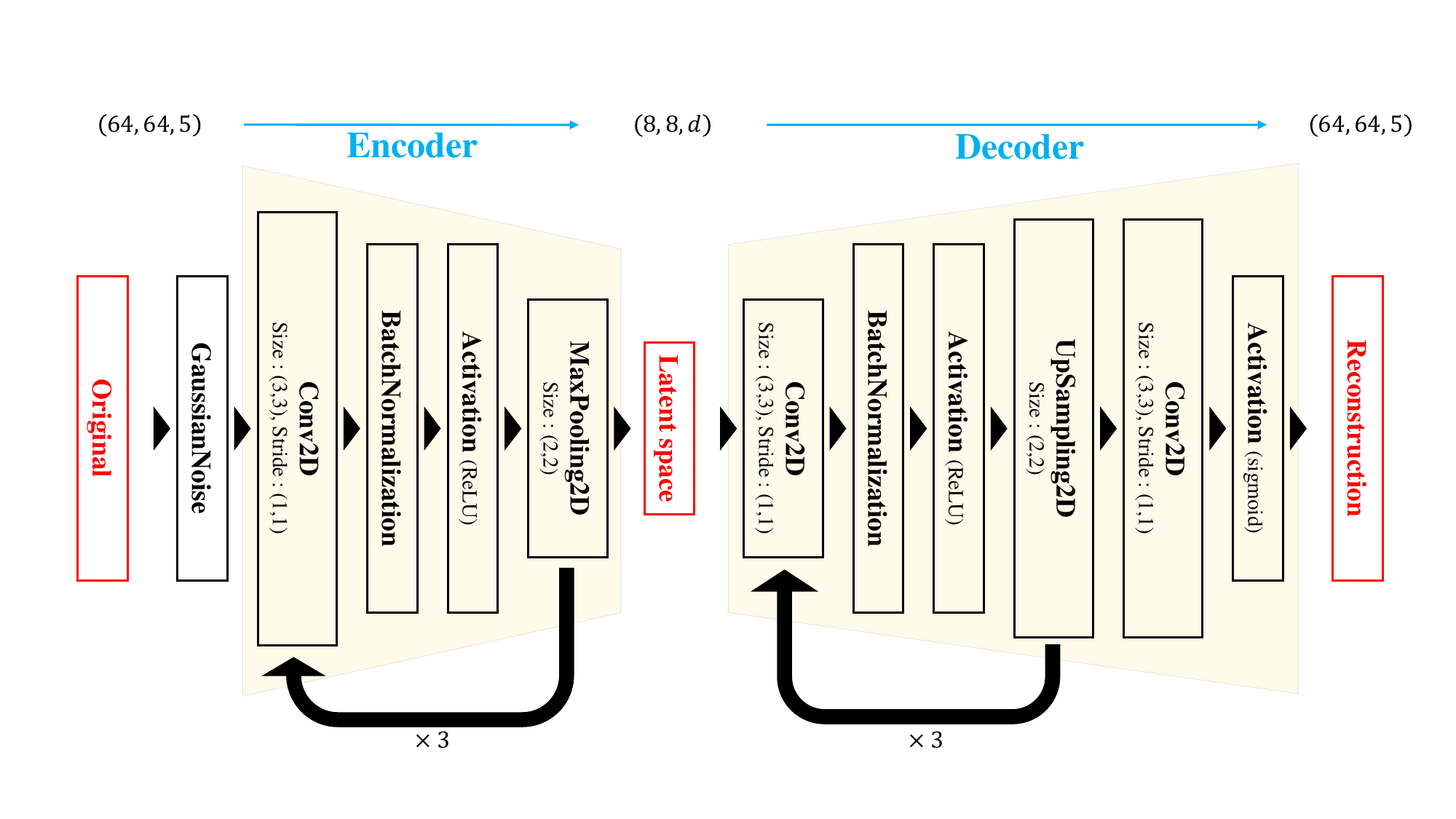} 
 \end{center}
\caption{
Architecture of our DCAE model. 
Input (original) data of a galaxy have $64\times64\times5$ dimensions. 
After adding Gaussian noise to the input data, the encoder produces data with reduced dimensions of $8\times8\times d$ in the latent space. 
The decoder processes these latent-space data to produce reconstruction data with the same dimensions as the input data. 
Each black-edged box indicates a layer in the model.
}
\label{model_image}
\end{figure*}

Here, we briefly explain how our DCAE model can serve as a platform for anomaly detection.
They trained a reproduction model with enormous data by updating the model weight parameters to minimize the loss function (see e.g., \cite{Chen2018,Zimmerer2018,Baur2021}).
Therefore, the trained model learns the overall structure representation of the training data, but not the uncommon features existing in them. 
Upon inputting a galaxy with anomalous features to the trained model, the model is unable to appropriately reconstruct those features and produce an image without them.
Consequently, the anomaly candidates can be selected by searching objects that manifest large residuals between the original and reconstructed images. 

\begin{figure*}
 \begin{center}
  \includegraphics[width=16cm]{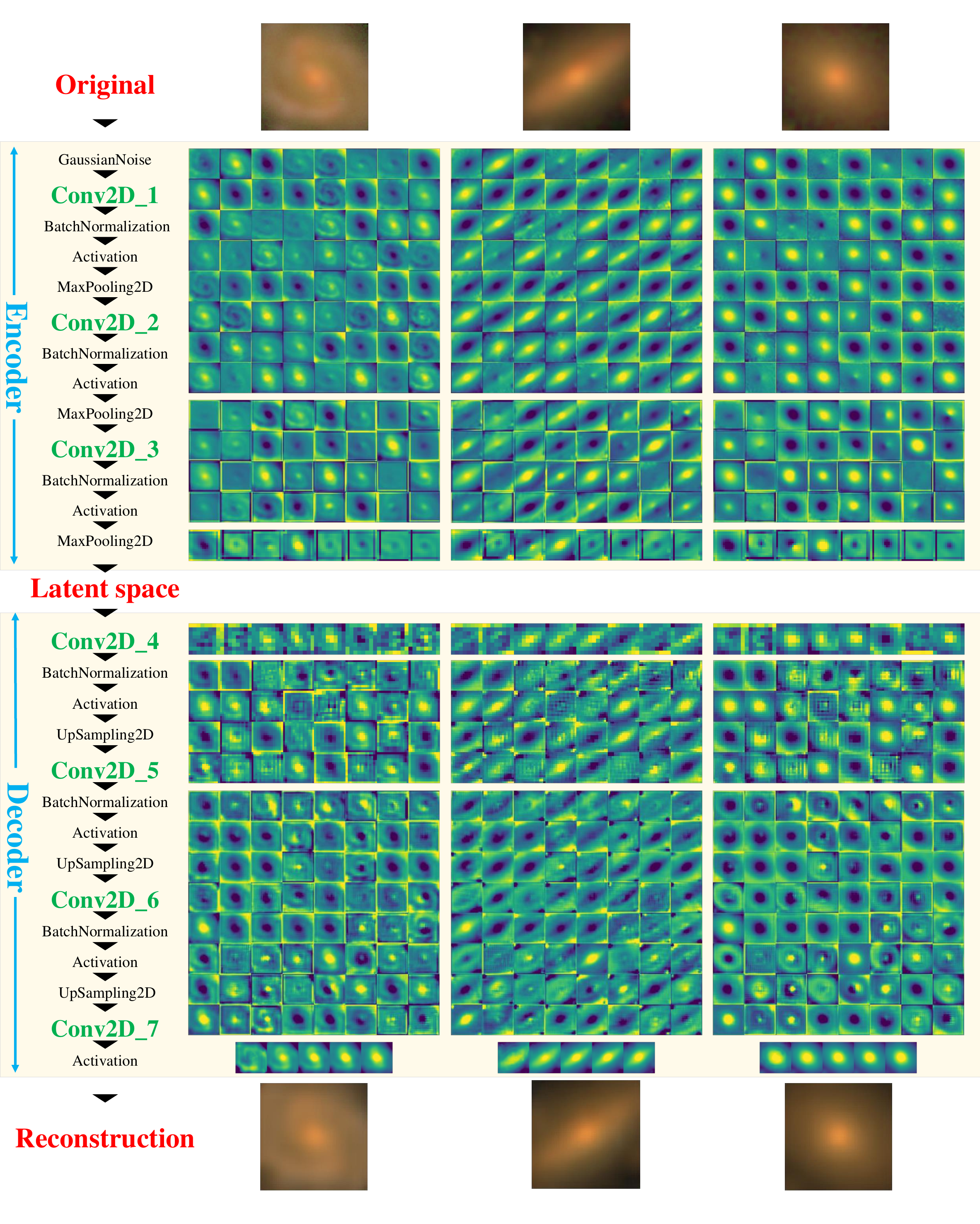} 
 \end{center}
\caption{
Outputs at each Conv2D layer for the three exemplary galaxies selected from the training sample 
(left: {\tt object\_id=40022257610803902} at $z=0.104$, 
middle: {\tt object\_id=40031324286770027} at $z=0.108$, 
right: {\tt object\_id=40031182552849262} at $z=0.105$). 
The top row presents the original image of each galaxy, and the second and third rows depict the outputs in individual Conv2D layers.
Conv2D\_1 to Conv2D\_3 are layers in the encoder, whereas Conv2D\_4 to Conv2D\_7 are layers in the decoder (refer to figure~\ref{model_image}).
The bottom row portrays reconstructed images, which have been normalized for ease of viewing.}
\label{activation_map}
\end{figure*}

In this study, we adopted {\tt TensorFlow} (version 2.5.0; \cite{Abadi2016}) and 
{\tt Keras} (version 2.5.0; \cite{Chollet2015}) to construct the DCAE model. 
In the training process, we leveraged the {\tt Google Colaboratoy} \citep{Bisong2019} to expedite the calculations with a GPU environment. 
The model architecture is illustrated in figure~\ref{model_image}.
The input images were five-color ($grizy$) images, each containing $64\times64$ pixels, as described in (section~\ref{s2}). 
The encoder contained convolutional layers (Conv2D), batch normalization layers (BatchNormalization; \cite{Ioffe2015}), activation function layers (activation), and max-pooling layers (MaxPooling2D). 
Convolutional layers spatially slide $3\times3$ filters over the images and produce feature maps, and then max-pooling layers take the max values in $2\times2$ windows. 
Batch normalization layers normalize the output of prior units (Conv2D in this study) by every batch, which increases training efficiency and prevents overfitting \citep{Ioffe2015}. 
Activation function layers convert outputs of prior units using an activation function that is usually nonlinear and enables complex representations throughout the NN (CNN).
The following rectified linear unit (ReLU; \cite{Glorot2011}) was adopted as the nonlinear activation function except the last layer using the sigmoid function:
\begin{linenomath}
\begin{equation}
    f(x) = \left\{ \begin{array}{ll}
    0 & \left(x\le0\right) , \\
    x & \left(x>0\right) .
  \end{array} \right.
  \label{ReLU}
\end{equation}
\end{linenomath}
Through Conv2D and MaxPooling2D layers in the Encoder, the input data dimensions are reduced to $8\times8\times d$ in the latent space (figures~\ref{model_image} and \ref{activation_map}), wherein $d$ denotes the number of channels in the final Conv2D layer within the encoder.
Thus, a smaller $d$ value provides stronger constraints on the latent representation. 
We selected $d=8$ as a trial-and-error process (appendix~\ref{A1}). 

On the other hand, the decoder consisted of Conv2D, BatchNormalization, Activation, and upsampling layers (UpSampling2D).
Upsamling layers simply resize the input images by a scale of two.
In addition, the ReLU was adopted as the activation function in the activation layers, except for the final layer in which a sigmoid function was applied:
\begin{linenomath}
\begin{equation}
    f\left(x\right) = \frac{1}{1+e^{-x}} \ \ .
\end{equation}
\end{linenomath}
The filter numbers in each Conv2D layer were 8, 32, 64, and 5, respectively (figure~\ref{activation_map}). 

The deep-learning model was trained using the training data prepared in section~ref{s2}, and the {\tt Adam} optimizer \citep{Kingma2014} was employed to adjust the weight parameters of the model and minimize the loss function value.
The optimizer updates the weight parameters using the backpropagation algorithm \citep{Rumelhart1986}. 
A learning rate is a scalar that determines how much weight parameters are updated in each step.
We trained the model with a learning rate of $l_r=0.001$ until a learning epoch of 10, after which it was altered to $l_r=0.0002$ (referred to as the learning decay technique in the machine-learning training). 
Figure~\ref{activation_map} explains how imaging data of spiral and early-type galaxies were encoded and decoded by our reproduction model. 
Additionally, the examples of reconstructed images and residuals in the five bands for normal galaxies from the training data, DR16Q quasars, and XELGs (as defined in subsection~\ref{s22}) are illustrated in figure~\ref{residuals}.
Although the proposed model appropriately reconstructed the normal galaxies, we observed significant residuals between the original and reconstructed images for most of the quasars and XELGs.
This is because the trained model was not optimized for uncommon color features originating from sources such as nuclear UV radiation in quasars and strong emission lines, as seen in XELGs.

\subsection{Selection of anomaly sources}\label{s32}

\begin{figure*}
 \begin{center}
  \includegraphics[width=15cm]{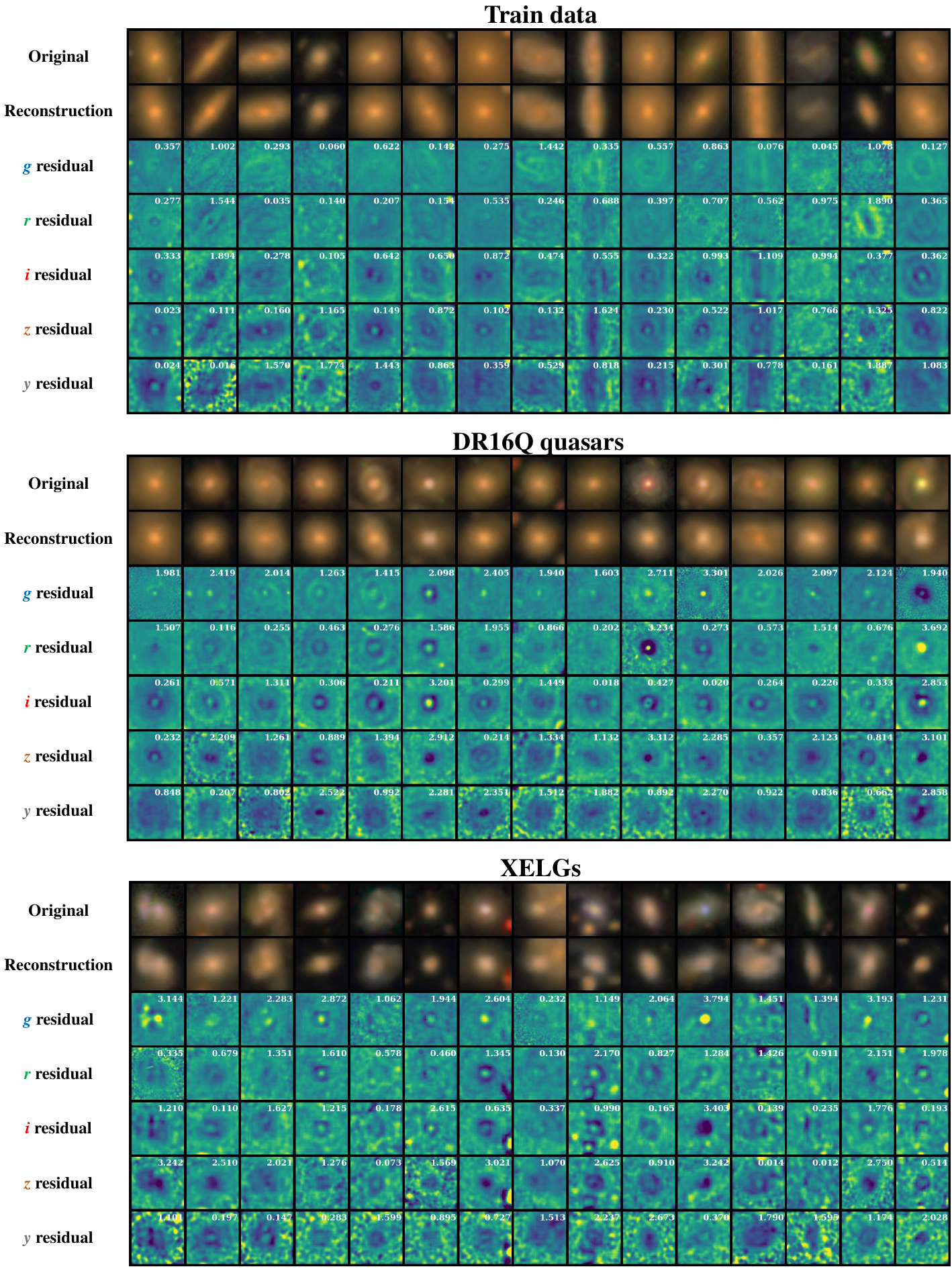} 
 \end{center}
\caption{
Original $gri$ data (top), reconstructed $gri$ data (second row), and residual data in each band (3rd--5th rows) for the randomly selected 15 galaxies in the three samples (top to bottom: training data, DR16Q quasars, and XELGs). 
The number presented in white in the top-right corner in each residual image indicates the $\sigma$ value of the $S_{\rm anom}$ excess calculated from the cumulative distribution function of $S_{\rm anom}$ for the training data.
}
\label{residuals}
\end{figure*}

To select objects with large residuals, such as quasars and XELGs, we defined the anomaly score $S_{\rm anom}$ as:
\begin{linenomath}
\begin{equation}
    S_{\rm anom}^{(\rm band)} = \sum_{{\rm k}\in A'} \left( \frac{\rm Source_{\rm \ k}^{(\rm band)} - \rm Reconstruction_{\rm\  k}^{(\rm band)}}{\rm Reconstruction_{\rm \ k}^{(\rm band)}} \right)^2,
\label{S_anom definition}
\end{equation}
\end{linenomath}
where $A'$ means a calculation area described below.
Thus, $S_{\rm anom}$ represents the MSE standardized by the square of the pixel counts of the reconstructed image, which gives the magnitude of the deviation from the model prediction.
In particular, the area $A'$ over which $S_{\rm anom}$ is calculated can be varied according to the target type of anomaly detection.
Because this study aimed to detect anomalies that have uncommon features in the nuclear regions such as quasars and some XELGs, we only used the central circular area with a diameter of 6 pixels as the calculation area $A'$, which broadly corresponds to the seeing FWHM (1 arcsec; subsection~\ref{s23}). 
Note that we may not be able to detect XELGs with their emission-line dominant components in the outer regions based on this definition.
The most suitable weight data was defined to enable the selection of maximum number of DR16Q quasars and XELGs based on a model with high selection completeness. 
The model selection procedure is detailed in appendix \ref{A1}. 
Sigma excess values of $S_{\rm anom}$ for quasars and XELGs are provided in figure~\ref{residuals}. The significant improvements in the anomaly scores in most of the DR16Q quasars and XELGs were compared to those of the normal galaxies derived from the training sample.  

\begin{figure*}
 \begin{center}
  \includegraphics[width=15cm]{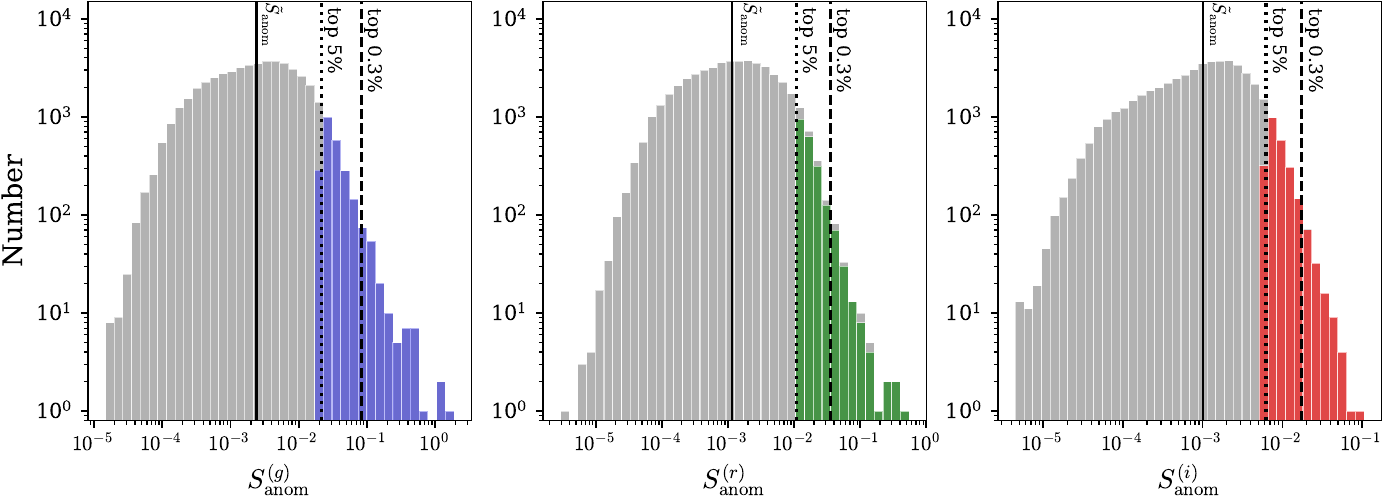} 
 \end{center}
\caption{
Distributions of the anomaly scores $S_{\rm anom}$ in three selection bands (left to right, the $g$, $r$, and $i$ bands) for the test data (gray) and the anomaly candidates (colored histograms). 
The solid, dotted, and dashed black lines depict the median $\tilde{S}_{\rm anom}$, top 5\% selection criteria, and top 0.3\% selection criteria in the corresponding selection bands, which correspond to $2\sigma$ and $3\sigma$ selection criteria respectively.
}
\label{sanom_hist}
\end{figure*}

\begin{figure*}
 \begin{center}
  \includegraphics[width=17cm]{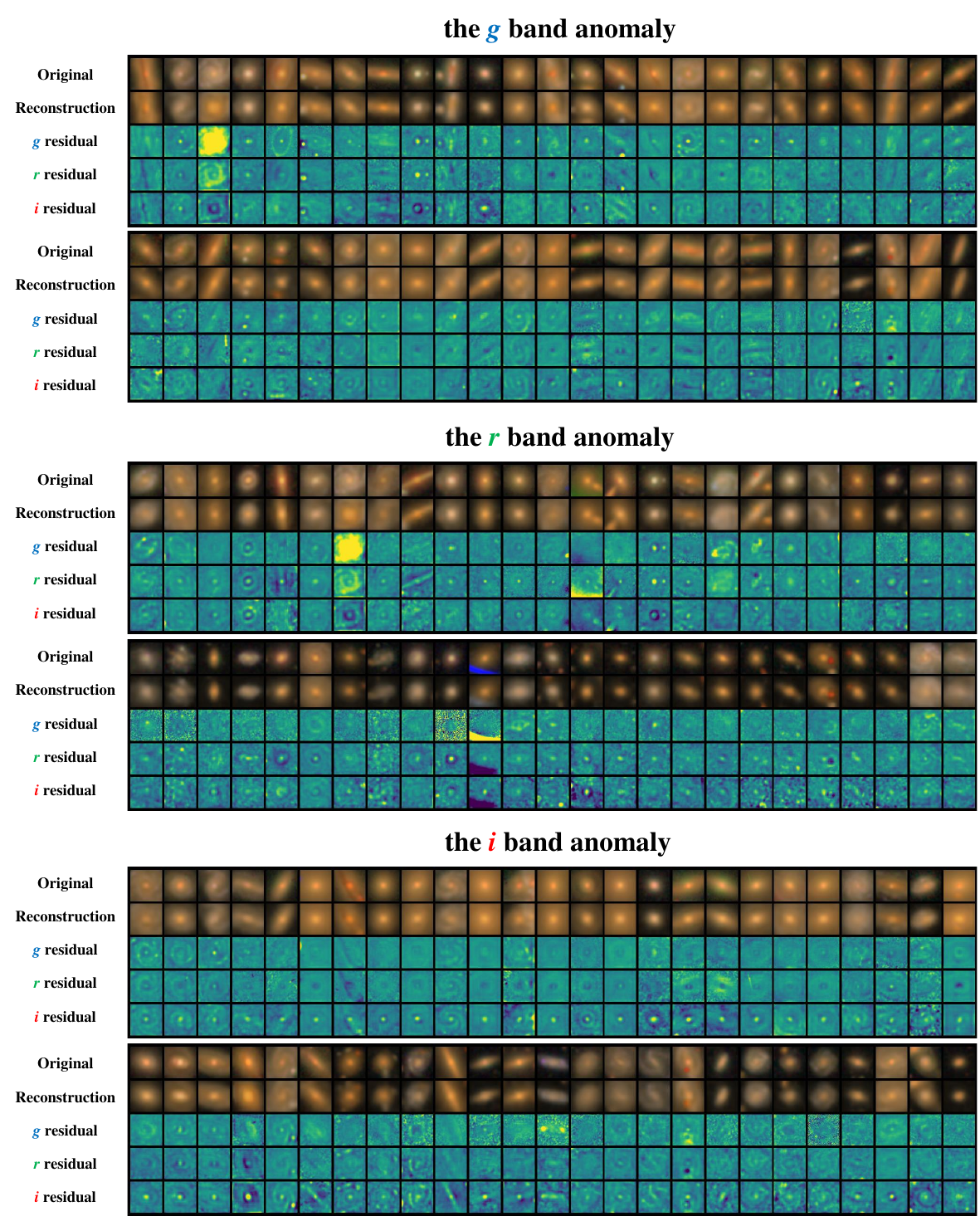} 
 \end{center}
\caption{
Original images (top row), reconstructed images (second row), and residual maps in the $g,r,i$ band (3rd--5th rows) of randomly selected 25 galaxies from the anomaly candidates selected in each band (top to bottom: the $g$, $r$ and $i$ bands).
}
\label{residuals_anomaly}
\end{figure*}

From the entire sample, we selected 5955 anomaly candidates by imposing the criteria: $S_{\rm anom}$ is in the top-5th percentile corresponding to the $2\sigma$ or higher excess (refer to figure~\ref{sanom_hist}), and the residual flux in the central area is positive in at least one of the $g$, $r$, and $i$ bands.
Their identification numbers ({\tt object\_id}), sky coordinates in the HSC-SSP PDR3 \citep{Aihara2021}, and anomaly scores are summarised in table~\ref{tab;catalog1}. 
Furthermore, examples of the selected anomaly candidates in each filter are represented in figure~\ref{residuals_anomaly}. 
The most anomalous galaxies in the current sample are discussed in subsection~\ref{s44}. 

We determined that 61\% of the DR16Q quasars $\left(=14 / 23\right)$ and 63.2\% of the XELGs $\left(=443/701\right)$ satisfy the above-mentioned criteria.
In contrast, there exist sources that have sufficiently high anomaly scores but are excluded from selection owing to their negative residual values, e.g., the numbers of those galaxies are 8, 322, and 14 for the $g$, $r$, and $i$ bands, respectively.
These objects are further discussed in appendix~\ref{A2}. 
Notably, as we evaluate the anomalies using only the central regions (equation~\ref{S_anom definition}), the chance projections of the background galaxies such as red galaxies will not affect the selection till they are not present in the central regions.

\begin{table*}
  \caption{Numbers of selected galaxies as anomalies}{%
  \begin{tabular}{rrlrlrl}
      \hline
      \multirow{2}{*}{band} & \multicolumn{2}{c}{Test data} & \multicolumn{2}{c}{DR16Q quasar} & \multicolumn{2}{c}{XELG}\\ 
           & \multicolumn{2}{c}{(N=49~319)} & \multicolumn{2}{c}{(N=23)} & \multicolumn{2}{c}{(N=701)}\\
      \hline
      $g$ & 2461 & (4.990\%) & 7 & (30\%) & 263 & (37.5\%) \\
      $r$ & 2147 & (4.353\%) & 3 & (13\%) & 119 & (17.0\%)\\
      $i$ & 2455 & (4.978\%) & 9 & (39\%) & 282 & (40.2\%) \\
      $g\,\&\,r$ & 391 & (0.793\%) & 1 & (4\%) & 77 & (11\%) \\
      $r\,\&\,i$ & 385 & (0.781\%) & 2 & (9\%) & 51 & (7.3\%) \\
      $g\,\&\,i$ & 418 & (0.848\%) & 2 & (9\%) & 119 & (17.0\%) \\
      $g\,\&\,r\,\&\,i$ & 86 & (0.17\%) & 0 & (0\%) & 26 & (3.7\%) \\
      \hline
      Total & 5955 & (12.07\%) & 14 & (61\%) & 443 & (63.2\%) \\
      \hline
    \end{tabular}}\label{tab;selnum}
\end{table*}

\begin{table*}
\caption{
List of identification numbers ({\tt object\_id}), coordinates, selection bands, and anomaly scores of the anomaly candidates from test data. 
See supplementary data 1 for the full version of this catalog (online material).
}
  \begin{tabular}{rrrrr}
      \hline
      ID & R.A. [deg] & Dec. [deg] & Band & $S_{\rm anom}^{(g)}, S_{\rm anom}^{(r)}, S_{\rm anom}^{(i)}$\\ 
      \hline
      36411461425187990 &  30.58684 & -5.99814 & $i$ & (0.00076, 0.00982, 0.00708)\\
      36411590274208581 &  30.44342 & -6.28742 & $g$ & (0.03182, 0.00485, 0.00031)\\
      36415992615686651 &  31.99153 & -6.09370 & $g$ & (0.05844, 0.00023, 0.00489)\\
      36416117169733290 &  31.81884 & -6.76016 & $i$ & (0.00296, 0.00096, 0.00786)\\
      36416409227512668 &  31.37596 & -5.98973 & $g$ & (0.05086, 0.00229, 0.00056)\\
      ... & ... & ... & ... & ...\\
      \hline
    \end{tabular}
\label{tab;catalog1}
\end{table*}

\section{Discussion}\label{s4}

In this section, we initially discuss the selection of quasars and XELGs in subsection~\ref{s41}.
Thereafter, the counterparts of the anomaly candidates were searched from the AGN catalogs, as described in subsection~\ref{s42}, and discuss the physical properties of anomaly candidates in subsection~\ref{s43}.
Lastly, we discuss the most rare and unique objects in the sample (subsection~\ref{s44}). 

\subsection{Validation with quasars and XELGs}\label{s41}

Our method selected 61\% of the DR16Q quasars (= 14/23) and 63.2\% of the XELGs (= 443/701). 
As these fractions are much higher than the anomaly fraction over the entire sample, 12.07\% (= 5955/49319), the proposed method can preferentially select quasars and XELGs.
To evaluate the performance of our anomaly detection method, we compared known quasars and XELGs selected as anomaly sources by the proposed method with the unselected known sources. 
We focused on radial profiles for quasars and emission-line EWs for XELGs because they are, respectively, the vital features characterizing these two populations.
This comparison will assist us in interpreting the kind of features that can be identified by the proposed machine-learning model. 

\begin{figure*}
 \begin{center}
  \includegraphics[width=17cm]{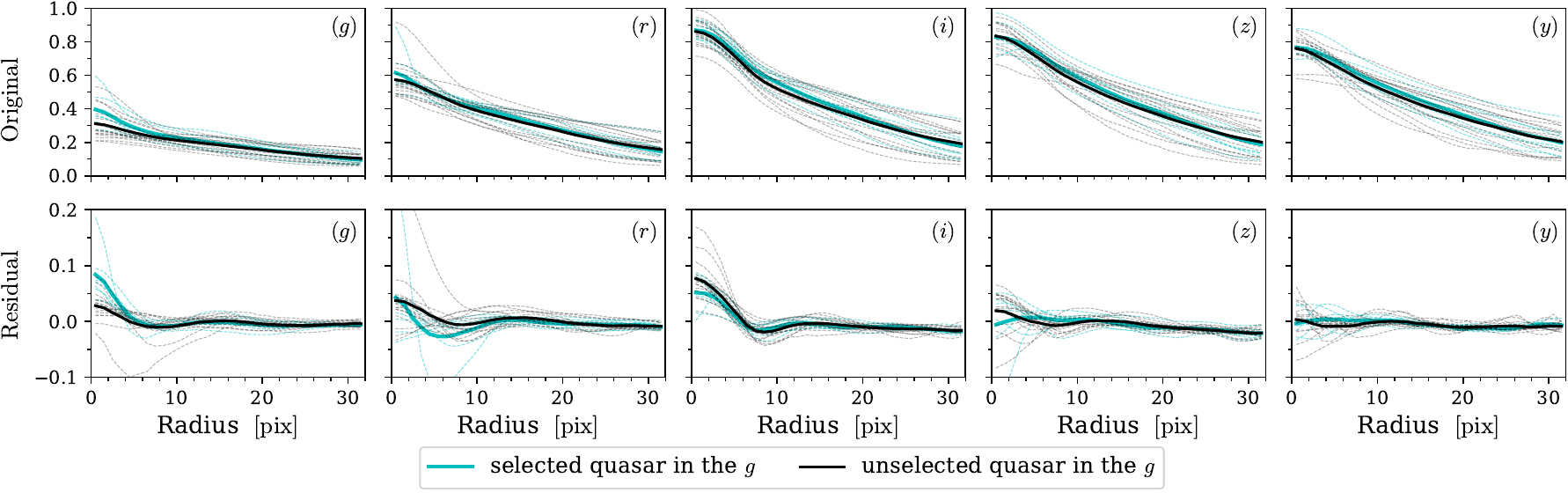} 
 \end{center}
\caption{
Comparison of radial profiles for selected and unselected DR16Q quasars in the $g$ band. 
The upper and lower rows depict the radial profiles of original and residual data, respectively, where the left to right columns denote $g,r,i,z,y$. 
Cyan and gray dashed lines individually  represent selected and unselected objects, respectively.
Cyan and black solid lines denote means of selected and unselected objects, respectively.
}
\label{DR16Q_comparison}
\end{figure*}

First, the quasars were examined. 
For ease of explanation, we focused on the quasars that are selected and not selected in a specific band, the $g$ band ($N=7$ and $16$, respectively; see table~\ref{tab;selnum}). 
The original and residual radial profiles for the quasars selected and not selected in the $g$ band are presented in figure~\ref{DR16Q_comparison}.  
As compared to the unselected objects, the selected objects exhibit significant nuclear excesses in the residual image.
Accordingly, we confirmed that the model can select quasars in cases of sufficiently large $g$ band excesses present in the center. 
As the model was not optimized for the unique color features observed in type-I AGNs and quasars, model reconstructions were generated without the blue (the $g$ band) nuclear components, as identified in the residual profiles (figure~\ref{DR16Q_comparison}).
However, the proposed model could accurately reproduce the $g$ band images of the unselected quasars in the $g$ band, which exhibited inadequate $g$ band excesses, regardless of the clear nuclear components in the $i$ band as well as the selected quasars. 
Note that the quasars and AGNs could be variable objects, and the SDSS spectra were obtained a decade ago. 
Therefore, certain known quasars could be inactive during the operation of the HSC-SSP (March 2014--January 2020 for HSC-SSP PDR3; \cite{Aihara2021}).

Furthermore, the XELGs were examined by focusing on the EWs of strong emission lines, as their intense emission-line contributions toward the selection band would yield high anomaly scores (see also subsection~\ref{s43} about the redshift dependence of the anomaly selection rate).
In addition, the parameters $EW_g$, $EW_r$, and $EW_i$ ($EW_{\rm band}$) were defined to quantify the emission-line contributions to the $g$, $r$, and $i$ bands, respectively. 
Herein, $EW_{\rm band}$ denotes the sum of the observed EWs of the strong emission lines in each band. 
In this calculation, we considered the actual throughput of the corresponding HSC filters \citep{Aihara2021}. 
Moreover, depending on the spectroscopic redshift of each source, the following emission lines provided in the MPA-JHU DR8 catalog \citep{Brinchmann2004,Kauffmann2003,Tremonti2004} were considered: 
[O{\sc ii}]$\lambda\lambda$3726,3729,
[Ne{\sc iii}]$\lambda$3869,
H$\beta$,
[O{\sc iii}]$\lambda\lambda$4959,5007,
H$\alpha$,
[N{\sc ii}]$\lambda$6548,6584, and
[S{\sc ii}]$\lambda\lambda$6717,6731. 

\begin{figure*}
 \begin{center}
  \includegraphics[width=13cm]{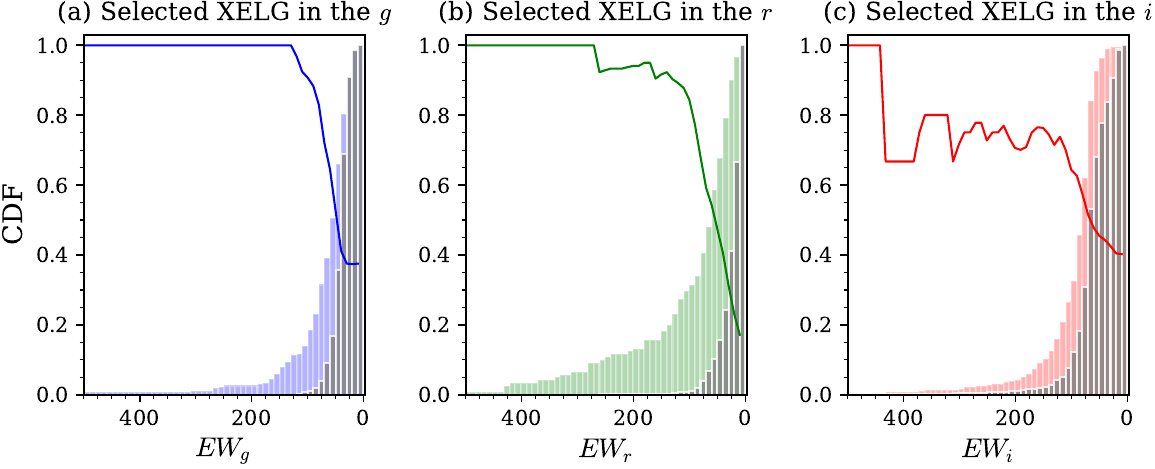} 
 \end{center}
\caption{
Cumulative EW distributions for selected (colored histograms) and unselected XELGs 
(black histograms) in the $g$ (left column), $r$ (middle), and $i$ (right) bands. 
The colored lines depict cumulative selection rate.
}
\label{XELG_comparison}
\end{figure*}

The cumulative distributions of $EW_{\rm band}$ for the selected and unselected XELGs are presented in figure~\ref{XELG_comparison}. 
The selection rates of objects with $EW_{\rm band}>100$ \AA\ was 90.7\% for the $g$ band, 84.4\% for the $r$ band, and 64.4\% for the $i$ band. 
Thus, the selection method can appropriately select the XELGs, given that they have prominent emission-line contributions to the target broadband images. 
More importantly, the high selection rate in the $g$ band could be caused by the $g$ band excess, which is a relatively uncommon trend in the training data.  

Based on these results, we concluded that the proposed model tends to select known rare objects with more anomalous features, specifically, quasars with a larger excess in the residual image and XELGs with intense emission lines. 
Note that the proposed model is not an identifier only for quasars and XELGs, and thus we do not discuss selection purity of these sources in this paper. 

\subsection{Evaluation with the Milliquas catalog }\label{s42}

In this subsection, the performance of the proposed model is determined based on the AGNs and quasars that were not used in the model training or selection process.
For this purpose, we cross-matched our anomaly source sample with the Milliquas (Million Quasars) Catalog v7.2 \citep{Flesch2021} that contains all the public AGNs and quasars as on April 30, 2021. 
Because we used the DR16Q sample in the model selection process, we excluded the DR16Q sample from the sample before matching with the Milliquas catalog.
AGNs and quasars in the Milliquas catalog matched with the test data are archived from the SDSS DR16 
\citep{Ahumada2020}, the low-redshift AGN catalog by \citet{Liu2019}, the 2dF galaxy 
survey \citep{Colless2001}, the 2dF QSO Redshift Survey \citep{Croom2004}, the 
Principal Galaxy Catalog \citep{Paturel2003}, the Large Sky Area Multi-Object Fiber 
Spectroscopic Telescope DR3/DR2 \citep{Dong2018}, the WISE AGN catalog 
\citep{Secrest2015}, and other papers 
\citep{Ge2012,Stalin2010,Francis2004,Croom2009,Lacy2007,Zhou2006,Gosset1997,Appenzeller1998,He1993,Markelijn1970,Bauer2000,Massaro2009,MacAlpine1981,Trichas2012,Trump2007,Lacy2013,Akiyama2015,Esquej2013,Garilli2014,Rowan2013,Foltz1989,Chang2019}.

The number of cross-matched sources within 1 arcsec apertures is summarized in table.~\ref{tab;milliquas}, where object categories follow the object types given in the Milliquas catalog: an optically core dominated object is labeled as “quasar" and a host dominated object is labeled as “AGN". 
In addition, \citet{Flesch2019} claimed that “Type-$\rm I\hspace{-.01em}I$ AGNs" may be contaminated by objects such as normal star-forming galaxies, LINERs, or starburst galaxies. 

\begin{table*}
  \caption{Results of cross-matching with the Milliquas catalog}
  \begin{tabular}{crrrrr}
      \hline
        & \multirow{2}{*}{Test data\footnotemark[$*$]} &\multicolumn{4}{c}{Selected anomalies\footnotemark[$*$]}\\
       &  & $g$ & $r$ & $i$ & total\\ 
       & (N=49~296) & (2454) & (2144) & (2446) & (5941) \\
      \hline
      quasar & 23 & 10 & 2 & 7 & 16 \\
      Type-$\rm\,I\,$ AGN & 286 & 62 & 20 & 87 & 130 \\
      Type-$\rm I\hspace{-.01em}I$ AGN & 580 & 35 & 11 & 40 & 74 \\
      Others\footnotemark[$\dag$] & 49 & 6 & 5 & 7 & 14 \\
      \hline
      total & 938 & 113 & 38 & 141 & 234 \\
      \hline
    \end{tabular}\label{tab;milliquas}
    \begin{tabnote}
    \footnotemark[$*$] 
    The DR16Q sample are excluded.\\
    \footnotemark[$\dag$]
    Objects with quasar-like features, such as BL Lac objects, or quasar candidates.\
    \end{tabnote}
\end{table*}

We determined that the proposed model selected 70\% $\left(=16 / 23\right)$ of the quasars and 45.5\% $\left(=130 / 286\right)$ of type-$\rm\,I\,$ AGNs, which are much higher than the anomaly fraction over the entire sample, 12.07\% $\left(=5955 / 49319\right)$.
Thus, these significant fractions suggested that the proposed method preferentially selected the quasar and type-$\rm\,I\,$ AGN samples.
On the contrary, the selection rate of type-$\rm I\hspace{-.01em}I$ AGNs was 12.8\% $\left(=74 / 580\right)$, which was as low as the anomaly fraction. 
Thus, the proposed method is not advantageous in searching type-$\rm I\hspace{-.01em}I$ AGNs. 
The lower selection rate of type-$\rm I\hspace{-.01em}I$ AGNs was potentially caused by their smaller impacts on host galaxies in comparison to type-$\rm I\hspace{-.01em}I$ AGNs in the optical regime. 
Thus, we concluded that an anomaly detection method, once optimized with appropriate test data, would be applicable for a blind search of quasars and type-$\rm\,I\,$ AGNs. 

Next, the bandpass dependence of the selection rate is discussed. 
For quasars, type-$\rm\,I\,$ AGNs, and type-$\rm I\hspace{-.01em}I$ AGNs, the selections based on the $g$ and $i$ bands yield three to five times higher selection rates than those based on the $r$ band (table~\ref{tab;milliquas}). 
Such high selection rates in the $g$ band would be attributed to luminous UV radiation from the accretion disk, while those in the $i$ band would be caused by strong H$\alpha$ and [N{\sc ii}] emission lines existing in this band.
For the remaining AGN class (“others"), the selection rate was similar among the three bands.
In contrast, the relatively lower selection rates in the $r$ band could be potentially because of their low $EW_r$ ($EW_r<20$ \AA).
Notably, the Milliquas catalog is incomplete, and therefore, the above-mentioned AGN fractions may increase upon considering the quasars or AGNs missing in the previous surveys.

\subsection{Characteristics of anomaly candidates}\label{s43}

Until the previous subsection, we determined that only 14 (from the DR16Q sample) and 234 (from the Milliquas catalog) sources ($\sim4\%$) among the 5955 anomaly candidates bear a counterpart in the currently available quasar and AGN catalogs. 
We performed complementary analyses to examine the characteristics of all the candidates using literature data in this subsection. 
This characterization will provide valuable insights toward future improvements of the proposed anomaly detection model.

\begin{figure*}
 \begin{center}
  \includegraphics[width=14cm]{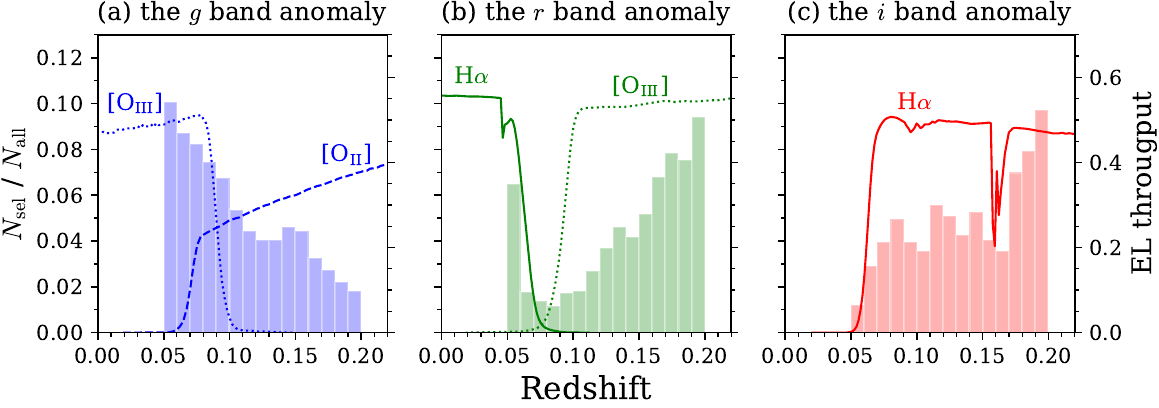} 
 \end{center}
\caption{
Selection rate as a function of redshift for the $g$ (left panel), $r$ (middle), and $i$ (right) bands. Solid, dashed, and dotted lines indicate the throughput of H$\alpha$, [O{\sc iii}], and [O{\sc ii}], respectively.
}
\label{sel_z}
\end{figure*}

Initially, we examined the redshift dependence of the selection rate shown in figure~\ref{sel_z}.  
The selection rate in a given band is defined as the ratio of the number of selected galaxies in that band, $N_{\rm sel}$, to the number of all galaxies in the test data, $N_{\rm all}$. 
As can be observed from figure~\ref{sel_z}a, the selection rate in the $g$ band decreases with the redshift.
This trend can be explained based on the behavior of the throughput of [O{\sc iii}] and [O{\sc ii}] emission lines with the redshift, i.e., the stronger [O{\sc iii}] line is captured by the $g$ band only at $z<0.09$, whereas the weaker [O{\sc ii}] line is within the band in most of the target redshift range.
Such a behavior can also explain an increase in the $r$ band selection rate at $z>0.09$. 
Similarly, a spike in the $r$ band selection rate at the lowest redshift and a relatively weak redshift bin dependence in the $i$ band selection rate are caused by the H$\alpha$ line. 
These results indicated that a significant fraction of the anomaly candidates were selected because of their strong emission line(s) and the selection rate in a given band is strongly dependent on what lines are redshifted into that band.

\begin{figure*}
 \begin{center}
  \includegraphics[width=14cm]{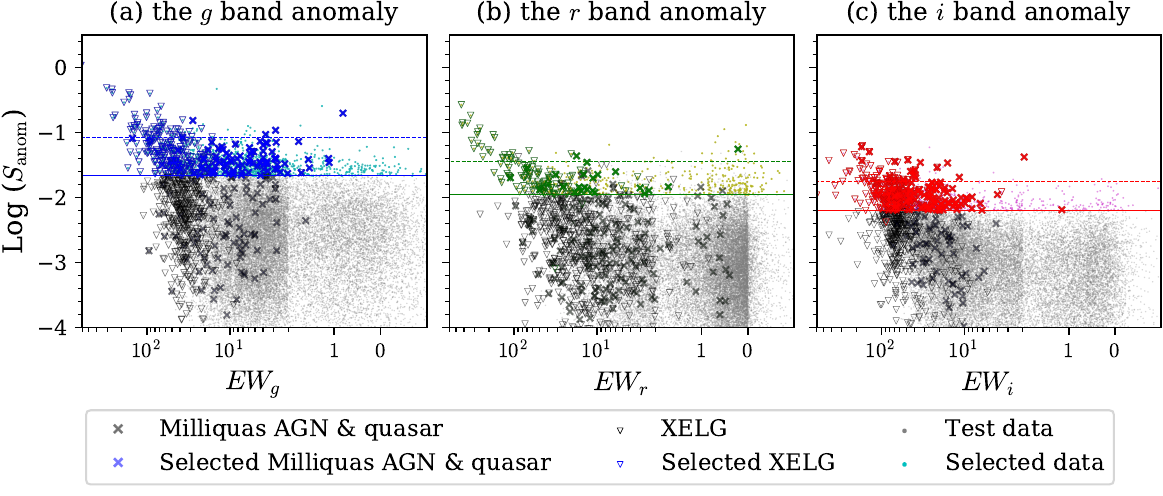} 
 \end{center}
\caption{
Relationship between $S_{\rm anom}$ and $EW_{\rm band}$ measured in the $g$ (panel [a]), $r$ ([b]), and $i$([c]) band.
The gray and colored dots denote all galaxies in test data and anomaly candidates selected in each band, respectively.
Similarly, the black and colored triangles represent all and selected XELGs, respectively, and the black and colored crosses indicate all and selected AGNs or quasars, respectively.
The solid, and dashed lines in each panel denote $S_{\rm anom}$ for the top 5\% ($2\sigma$) and top 0.3\% ($3\sigma$) selection criteria in the anomaly selection process, respectively (see subsection~\ref{s32}).
}
\label{EW_Sanom}
\end{figure*}

The influences of $EW_{\rm band}$ on the selection process were further investigated by directly comparing the $EW_{\rm band}$ with the anomaly score $S_{\rm anom}$ for 25~714 galaxies that have EW measurements in the MPA-JHU catalog in figure~\ref{EW_Sanom}, where the XELG sample is included as well. 
Again, we found that galaxies with high $EW_{\rm band}$ tend to exhibit high $S_{\rm anom}$, especially at $EW_{\rm band}>50$ \AA, which indicated that the proposed model preferentially selected the extreme emission-line objects with high anomaly scores. 
However, figure~\ref{EW_Sanom} depicts certain intriguing sources with high $S_{\rm anom}$ but low $EW_{\rm band}$ ($EW_{\rm band}<1$ \AA). 
More than half of these outlier objects apparently exhibited artificial-like features, as seen in appendix~\ref{A2}, although we cannot expunge the possibility that they have intense line-emission only at the center, which cannot be judged by one-dimensional SDSS spectra captured with the $3"$-aperture fibers owing to dilution.
Moreover, figure~\ref{EW_Sanom} indicates certain galaxies with high $EW_{\rm band}$ ($EW_{\rm band}>100$ \AA) but low $S_{\rm anom}$, especially in the $i$ band (see also figure~\ref{XELG_comparison}).
However, 70\% of these galaxies were selected in other band(s) as well, which implied that their low $S_{\rm anom}$ values can be caused by the flux excess in other bands. 
For instance, XELGs with low $S_{\rm anom,i}$ at $z=$ 0.1--0.2 can be explained based on the intense [O{\sc iii}] emission in the $r$ band. 
In such a case, the model may confuse XELGs with red galaxies and thereafter, reproduce their $i$ band excesses better than expected.  

\begin{figure*}
 \begin{center}
  \includegraphics[width=14cm]{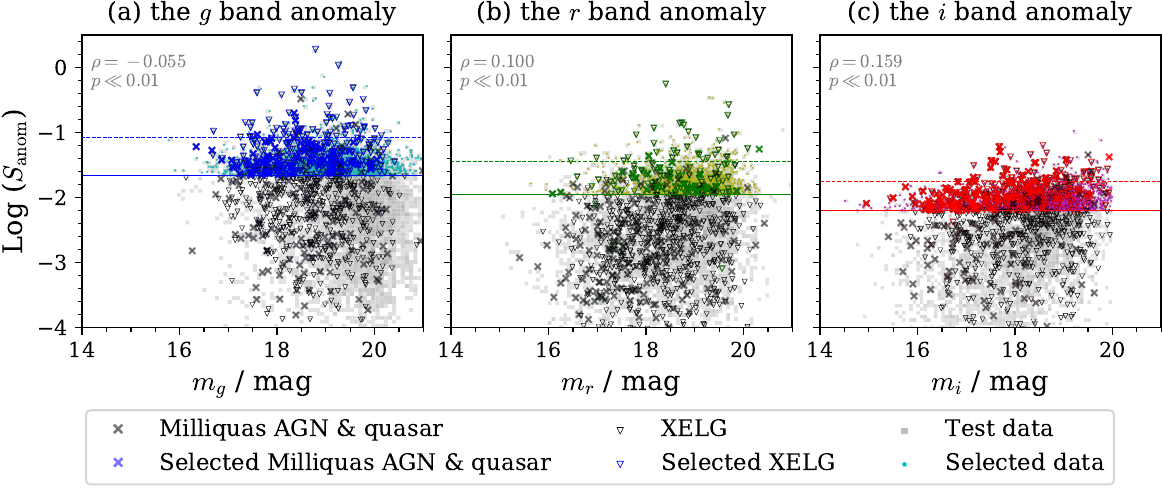} 
 \end{center}
\caption{
Relationship between $S_{\rm anom}$ and apparent magnitudes
in the $g$ (panel [a]), $r$ ([b]), and $i$([c]). The symbols and lines represent the same as that in figure~\ref{EW_Sanom}.
The Spearman's correlation coefficient and its $p$-value for galaxies in test data are depicted in the top-left corner of each panel.
}
\label{mag_Sanom}
\end{figure*}

Furthermore, the dependence of the anomaly score on apparent magnitude was examined. 
As $S_{\rm anom}$ denotes the square sum of the normalized residuals, it is expected not to be sensitively dependent on apparent magnitude, except for the faint sources  bearing low $S/N$ ratios per pixel.
The anomaly score $S_{\rm anom}$ is illustrated as a function of apparent magnitude for the entire sample in figure~\ref{mag_Sanom}.
The top 5\% and 0.7\% (correspond to $2\sigma$ and $3\sigma$) criteria of $S_{\rm anom}$ for the respective bands are depicted by solid and dashed lines, respectively. 
In addition, Spearman's correlation test could not derive any correlation in the three bands, which assures that the proposed method can evaluate anomalies independently of the apparent magnitude in the sample.
Note that the anomalies selected in the $r$ band tend to be biased toward fainter magnitudes at $m_r>16.5$ because the luminous objects with $m_r<16.5$ and large $S_{\rm anom}$ tend to exhibit negative residuals. 
These sources are further examined in appendix~\ref{A2}. 

\begin{figure*}
 \begin{center}
  \includegraphics[width=17cm]{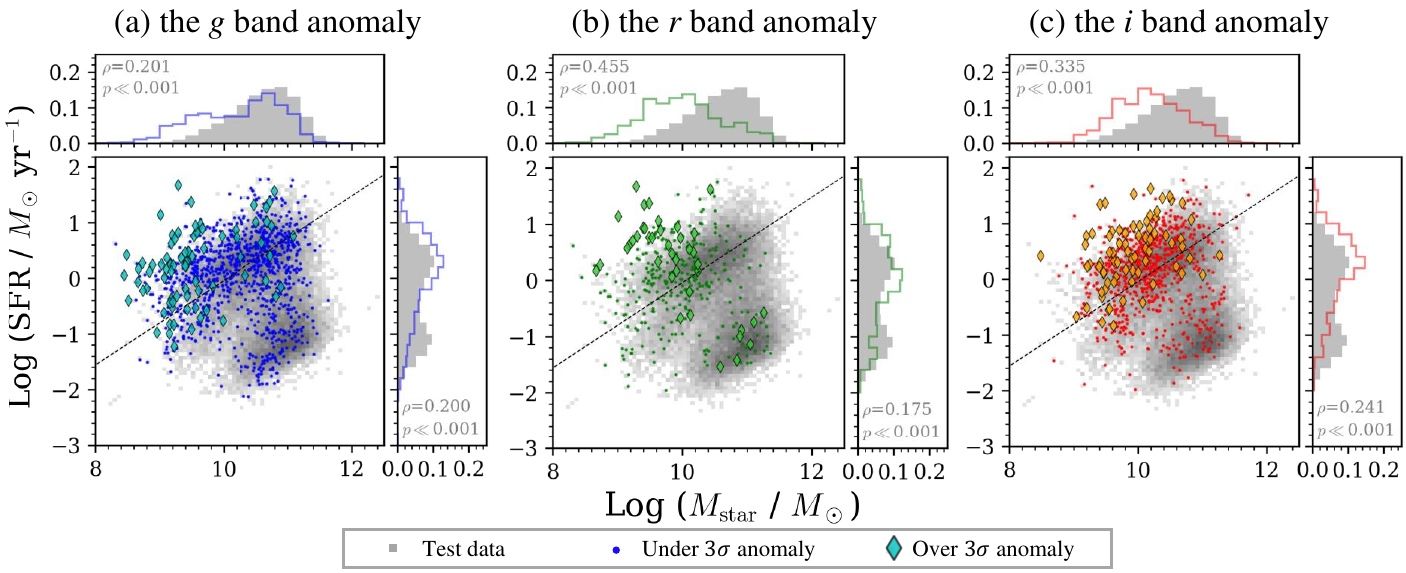} 
 \end{center}
\caption{
Relationship between the SFR and $M_{\rm star}$ for the sources selected in the $g$ (panel [a]), $r$ ([b]), and $i$ ([c]). 
For each panel, the gray and colored dots in the main plot denote galaxies in the test data and selected anomaly sources, respectively; diamond shapes indicate sources with $S_{anom}$ exceeding the $3\sigma$ upper limit.
Black dashed lines indicate the SFR-$M_{\rm star}$ relation for the star-forming main sequence relation from \citet{Renzini2015}}
The gray and colored histograms in the subplots represent the histograms of $M_{\rm star}$ and SFR for all anomaly sources, respectively.
KS test statistics $\rho$ and its $p$-value are shown in the top and right histograms.
\label{SFR_LGM}
\end{figure*}

Figure~\ref{SFR_LGM} indicates the distribution of the galaxies in the MPA-JHU catalog on the SFR--$M_{\rm star}$ plane. 
In figure~\ref{SFR_LGM}, the upper and lower crowds of objects are known to be the star-forming (SF) main sequence and the passive sequence \citep{Brinchmann2004}, respectively. 
Note that the quasars and AGNs are not plotted because their SFR and $M_{\rm star}$ would exhibit large uncertainties (e.g., \cite{Toba2018, Toba2021}).
In all the bands, the proposed model preferentially selects the galaxies in or above the SF main sequence, as indicated by the adequately small $p$-values ($\ll0.01$) in the Kolmogorov--Smirnov (KS) test. 
Thus, the result is broadly consistent with the trend that the current method preferentially selected the strong emission-line galaxies. 
We also confirmed a similar trend as in the GALEX-SDSS-WISE Legacy Catalog 2 (GSWLC-2: \cite{Salim2016,Salim2018}), which offers more reliable SFR and $M_{\rm star}$ measurements than the MPA-JHU catalog because of its multi wavelength coverage from UV to mid-IR. 
However, significant fractions of anomaly candidates (e.g. 78\% in the $r$ band selection) that do not appear in this case owing to the absence of counterparts and failures of the SED fit in the GSWLC-2 (see also subsection~\ref{s44}), and thus we decided not to use the measurements by the GSWLC-2 in this paper. 

\subsection{The most anomalous sources}\label{s44}

\begin{figure*}
 \begin{center}
  \includegraphics[width=17cm]{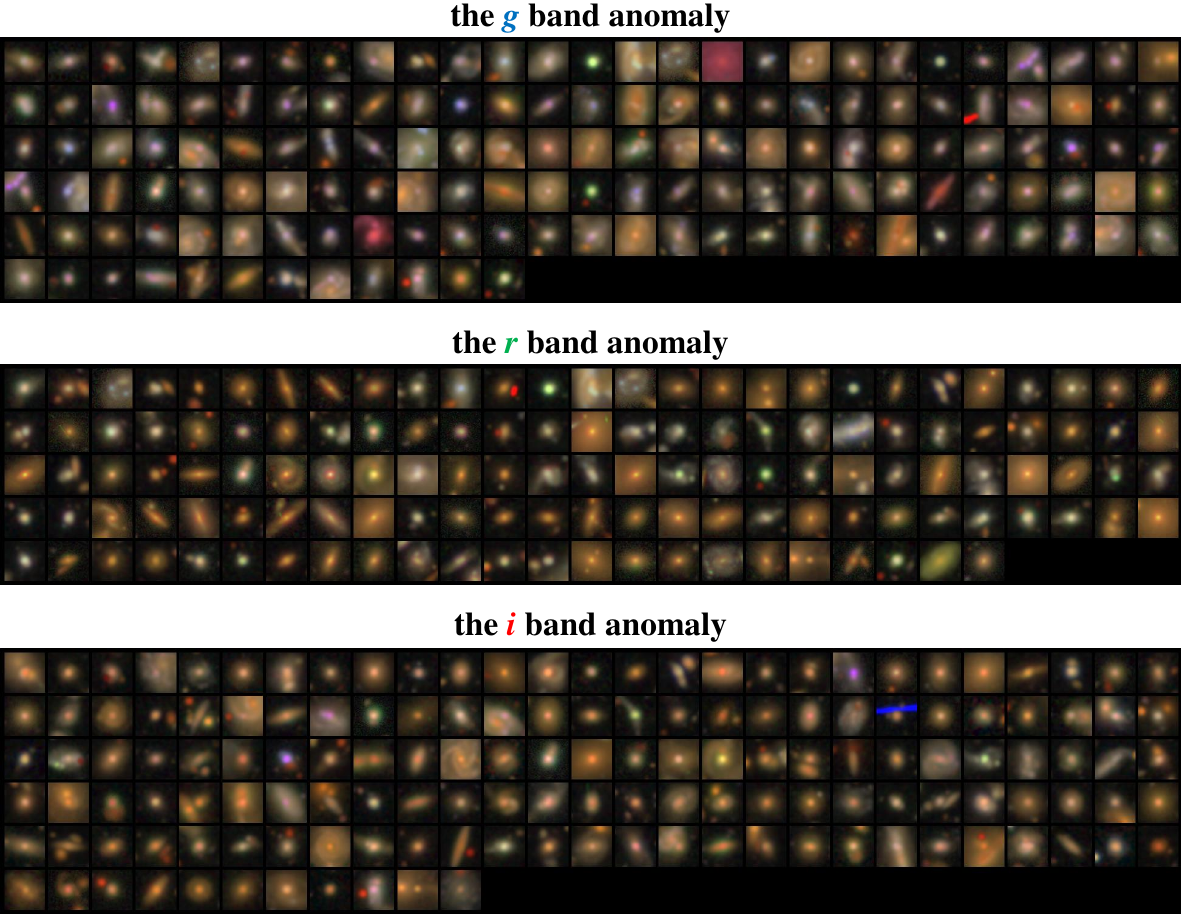} 
 \end{center}
\caption{
Postage stamps of anomaly candidates with top 
$0.3\%$ $S_{\rm anom}$ (exceeding $3\sigma$ upper limit) in the $g,r,$ and $i$ bands.
}
\label{anomaly_3sigma}
\end{figure*}

\begin{figure*}
 \begin{center}
  \includegraphics[width=14cm]{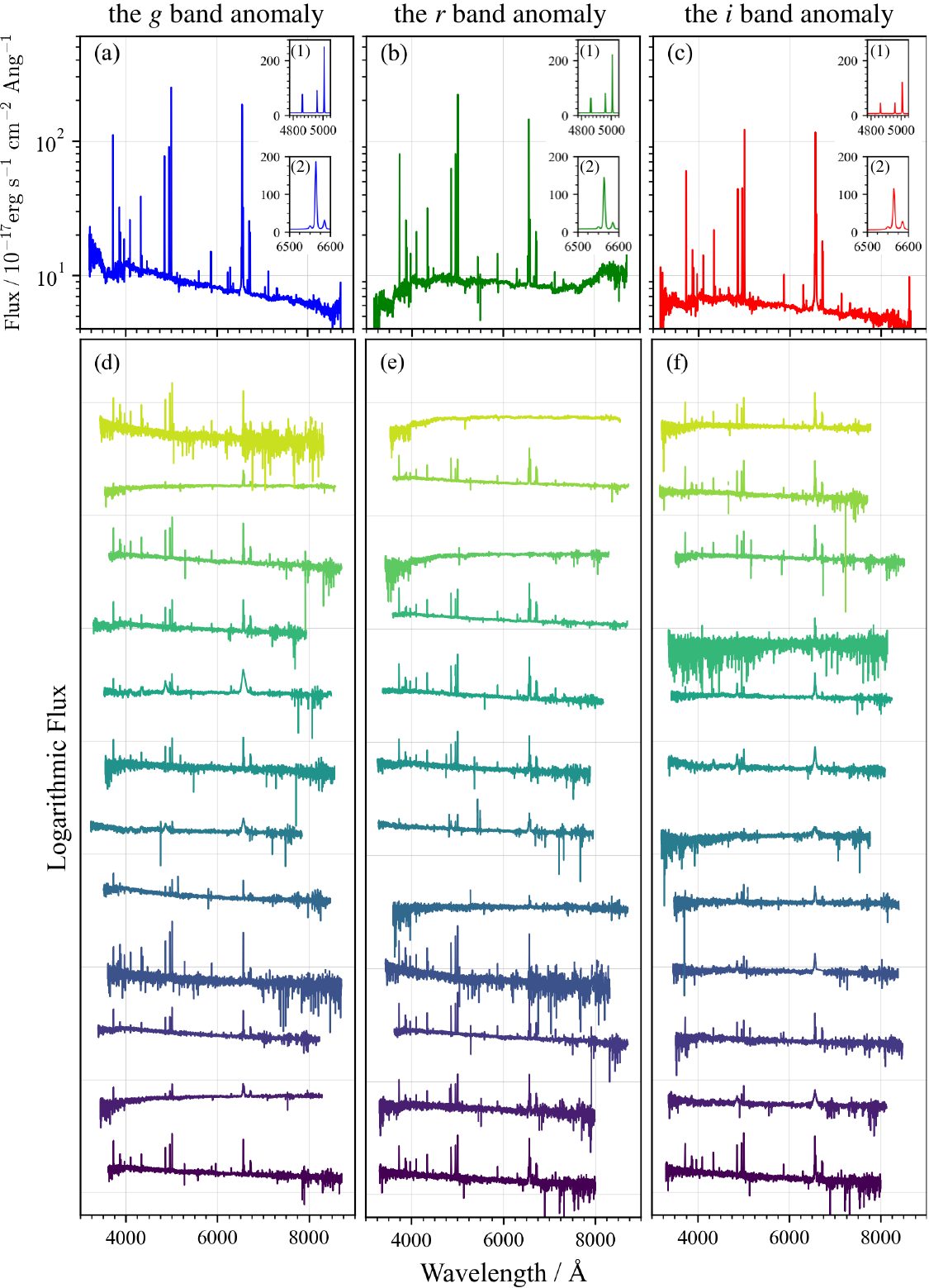} 
 \end{center}
\caption{
Upper panels (a), (b), and (c) indicate stacked spectra of all the anomaly candidates with the top $0.3\%$ $S_{\rm anom}$ (exceeds $3\sigma$ upper limit) in the $g,r,$ and $i$ bands, respectively, from the SDSS DR16 \citep{Ahumada2020}. 
Magnified views around (1) H$\beta$ and [O{\sc iii}]$\lambda\lambda$4959,5007, and (2) H$\alpha$, [N{\sc ii}]$\lambda$6548,6584 are shown in the right small graphs in an each panel. 
Lower panels (d), (e), and (f) indicate 12 examples of the spectra of anomaly candidates with the top $0.3\%$ $S_{\rm anom}$ (exceeds $3\sigma$ upper limit) in the $g,r,$ and $i$ bands, respectively, from the SDSS DR16 \citep{Ahumada2020}.
Each colored lines indicate common logarithm of spectrum for each object, and colors are changed in each panel for easy identification.
Each individual spectrum is shifted 
in the flux direction from one another.
}
\label{each_spec}
\end{figure*}

Ultimately, we examined the emission-line strengths, morphology, and star-formation activities of “the most anomalous sources", which are defined as with the $S_{\rm anom}$ higher than the $3\sigma$ (top $0.3\%$).
The dashed lines in figures~\ref{sanom_hist}, \ref{EW_Sanom}, and \ref{mag_Sanom} indicate an excess of $3\sigma$. 
In addition, figures~\ref{anomaly_3sigma} and \ref{each_spec} depict the original multicolored images and the SDSS DR16 \citep{Ahumada2020} spectra of the most anomalous sources, respectively. 

First, the XELGs were determined as dominant in the most anomalous sources.
In particular, 71\%, 48\%, and 65\% of the objects over the 3$\sigma$ excess from the MPA-JHU catalog were XELGs in the $g$, $r$, and $i$ bands, respectively.  
These results indicated that the proposed anomaly selection method with the $3\sigma$ excess, especially in the $g$ and $i$ bands, facilitated the blind searches of XELGs. 
However, as certain objects over the 3$\sigma$ excess in the $r$ band exhibited a small $EW_r$ ($EW_r<1$ \AA; figure~\ref{EW_Sanom}), the XELG fraction in the $r$ band was relatively low.

As can be observed in figure~\ref{anomaly_3sigma}, a number of the most anomalous sources exhibited merger-like features or close neighbors. 
In the former case, we confirmed that the residual images of such galaxies tend to have multiple components even in their residual images, suggesting that they are probably dual AGNs or quasars or have multi-emission-line dominant components. 
As observed in figure~\ref{anomaly_3sigma}, some of the most anomalous sources selected in the $g$ band portraying a blue component, such as the blueberry galaxies, also known as XELGs \citep{Yang2017}.
Similar trends can be observed in the $r$ or $i$ bands, as they selected galaxies with a green or red component, respectively. 
In addition, certain sources with a purple component were caused by the strong emission-line contributions from the [O{\sc iii}] (or [O{\sc ii}]) and H$\alpha$ to the $g$ and $i$ bands, respectively. 
Such intense emission-line features can be confirmed in their optical spectra, as depicted in figure~\ref{each_spec}.

The colored diamond plots in figure~\ref{SFR_LGM} illustrate the relationship between the SFR and $M_{\rm star}$ for the most anomalous sources. 
In all the band, the majority of the most anomalous sources are located above the SF main sequence.
Besides, the most anomalous sources tend to be less massive compared to the entire sample. 
Although some of the most anomalous sources in the $r$ band are located in the passive sequence, in appearance they are the artifact-like sources as discussed in appendix~\ref{A2}. 

We also comment about fractions of the invalid SFR--$M_{\rm star}$ measurements in the GSWLC-2 for the most anomalous sources (see subsection~\ref{s43}). 
We noticed that $\sim16$\% of the most anomalous sources cannot be appropriately fitted with the SED templates in the GSWLC-2 \citep{Salim2018}, which are remarkably higher than that of the complete test data, 0.67\%. 
Such a high fraction suggests that the proposed method selected anomalies that exhibited too anomalous photometric colors to apply the SED fitting, owing to extreme emission lines or emission from SMBH, as observed in figure~\ref{each_spec}. 
Therefore, specialized templates are required to appropriately derive these physical parameters, which constitute as the in-depth analyses in future research. 

Conclusively, the present anomaly candidate sample contained unique populations. 
Further investigations of such sources will help in the search for interesting objects such as dual quasars, XELGs, or galaxies with unusual merger-like features.
Thus, we expect that the application of the proposed deep-anomaly detection method on unprocessed larger data-sets can result in the discovery of extremely unique objects, including unknown unknowns.
This aspect was demonstrated by applying the current model on a magnitude-limited color-selected sample without spectroscopic counterparts, as described in appendix~\ref{A3}.

\section{Conclusion}\label{s5}

This paper proposed a method based on machine learning to select galaxies with an anomaly in their central region. Subsequently, the model was applied on 49~319 sources at $0.05<z<0.2$ in the HSC-SSP multiband data.
This method evaluated the degree of anomalies using a newly defined parameter, the anomaly score, which is a measure of the light excess in the central region relative to normal galaxies. 
Overall, we selected 5955 anomaly candidates. 
As $\sim 60-70\%$ of the quasars, $\sim 45\%$ of the type-I AGNs, and $\sim 60\%$ of the XELGs in the data-set were selected using this method, new quasars and AGNs that are still unidentified in prior surveys could be present in the remaining candidates.

In addition, the characteristics of the anomaly candidates were discussed to investigate the types of sources selected by the anomaly detection model. 
For the quasar sample, the selected quasars tended to display larger nuclear residuals in the selection band in comparison to the unselected quasars. 
Moreover, we determined that galaxies with larger $EW_{\rm band}$ tend to exhibit larger anomaly scores, which indicates that the emission-line dominant sources were preferably selected as anomaly candidates. 
Moreover, the most anomalous sources exhibiting the highest anomaly scores tended to 
display very high $EW_{\rm band}$, and 50$\sim$70\% of them were XELGs.
In addition, a number of the most anomalous sources depicted a merger-like feature or were close neighbors.
These results demonstrated the superior capability of the proposed anomaly detection method to search for unique populations in the universe.

One limitation of this study is the purity of the training sample.
Although the training data should ideally not contain anomalies; however, the training data may have undetected anomaly objects or unknown unknowns because these selection criteria depended on only the SDSS 1-D spectral data. 
A large survey with integral field units like the Mapping Nearby Galaxies at APO (MaNGA;  \cite{Bundy2015}) at $z\sim0.03$ may help us restrict and purify the training data in the future. 
Another solution would be the use of all data as training data without the cleaning by developing our model, which will enable a completely blind anomaly survey. 

The major advantage of the proposed anomaly detection method over ordinary methods for searching rare objects is that the current method does not require any model template of targets in advance. 
Furthermore, the current anomaly detection method is not a classifier or a detector dedicated to specific species such as quasars or XELGs, and therefore, can discover other unique objects, including unknown unknowns. 
In addition, the proposed model can be flexibly modified to detect anomalies other than those in central regions by adjusting the definition of the anomaly score.
Moreover, the application of the anomaly detection model on future large surveys, for instance, PSF-SSP \citep{Takada2014,Tamura2016}, LSST \citep{Ivezic2019}, Euclid \citep{Amendola2018}, and Roman Space Telescope \citep{Akeson2019} will provide an excellent opportunity to discover new astronomical sources. 


\begin{ack}
This work is based on data collected at the Subaru Telescope and retrieved from the HSC data archive system, which is operated by Subaru Telescope and Astronomy Data Center at National Astronomical Observatory of Japan. We are honored and grateful for the opportunity of observing the Universe from Maunakea, which has the cultural, historical and natural significance in Hawaii. 

We thank anonymous referee for helpful feedback.
This work was partially supported by the Summer Student Program (2020) by the National Astronomical Observatory of Japan and the Department of Astronomical Science, The Graduate University for Advanced Studies, SOKENDAI.

The Hyper Suprime-Cam (HSC) collaboration includes the astronomical communities of Japan and Taiwan, and Princeton University. The HSC instrumentation and software were developed by the National Astronomical Observatory of Japan (NAOJ), the Kavli Institute for the Physics and Mathematics of the Universe (Kavli IPMU), the University of Tokyo, the High Energy Accelerator Research Organization (KEK), the Academia Sinica Institute for Astronomy and Astrophysics in Taiwan (ASIAA), and Princeton University. Funding was contributed by the FIRST program from Japanese Cabinet Office, the Ministry of Education, Culture, Sports, Science and Technology (MEXT), the Japan Society for the Promotion of Science (JSPS), Japan Science and Technology Agency (JST), the Toray Science Foundation, NAOJ, Kavli IPMU, KEK, ASIAA, and Princeton University. 

This paper makes use of software developed for the Large Synoptic Survey Telescope. We thank the LSST Project for making their code available as free software at  \url{http://dm.lsst.org}

The Pan-STARRS1 Surveys (PS1) have been made possible through contributions of the Institute for Astronomy, the University of Hawaii, the Pan-STARRS Project Office, the Max-Planck Society and its participating institutes, the Max Planck Institute for Astronomy, Heidelberg and the Max Planck Institute for Extraterrestrial Physics, Garching, The Johns Hopkins University, Durham University, the University of Edinburgh, Queen’s University Belfast, the Harvard-Smithsonian Center for Astrophysics, the Las Cumbres Observatory Global Telescope Network Incorporated, the National Central University of Taiwan, the Space Telescope Science Institute, the National Aeronautics and Space Administration under Grant No. NNX08AR22G issued through the Planetary Science Division of the NASA Science Mission Directorate, the National Science Foundation under Grant No. AST-1238877, the University of Maryland, and Eotvos Lorand University (ELTE) and the Los Alamos National Laboratory.

Funding for the Sloan Digital Sky Survey IV has been provided by the Alfred P. Sloan Foundation, the U.S. Department of Energy Office of Science, and the Participating Institutions. SDSS-IV acknowledges support and resources from the Center for High Performance Computing  at the University of Utah. The SDSS website is \url{www.sdss.org}.

SDSS-IV is managed by the Astrophysical Research Consortium for the Participating Institutions of the SDSS Collaboration including the Brazilian Participation Group, the Carnegie Institution for Science, Carnegie Mellon University, Center for Astrophysics | Harvard \& Smithsonian, the Chilean Participation Group, the French Participation Group, Instituto de Astrof\'isica de Canarias, The Johns Hopkins University, Kavli Institute for the Physics and Mathematics of the Universe (IPMU) / University of Tokyo, the Korean Participation Group, Lawrence Berkeley National Laboratory, Leibniz Institut f\"ur Astrophysik Potsdam (AIP),  Max-Planck-Institut f\"ur Astronomie (MPIA Heidelberg), Max-Planck-Institut f\"ur Astrophysik (MPA Garching), Max-Planck-Institut f\"ur Extraterrestrische Physik (MPE), National Astronomical Observatories of China, New Mexico State University, New York University, University of Notre Dame, Observat\'ario Nacional / MCTI, The Ohio State University, Pennsylvania State University, Shanghai Astronomical Observatory, United Kingdom Participation Group, Universidad Nacional Aut\'onoma de M\'exico, University of Arizona, University of Colorado Boulder, University of Oxford, University of Portsmouth, University of Utah, University of Virginia, University of Washington, University of Wisconsin, Vanderbilt University, and Yale University.

GAMA is a joint European-Australasian project based around a spectroscopic campaign using the Anglo-Australian Telescope. The GAMA input catalogue is based on data taken from the Sloan Digital Sky Survey and the UKIRT Infrared Deep Sky Survey. Complementary imaging of the GAMA regions is being obtained by a number of independent survey programmes including GALEX MIS, VST KiDS, VISTA VIKING, WISE, Herschel-ATLAS, GMRT and ASKAP providing UV to radio coverage. GAMA is funded by the STFC (UK), the ARC (Australia), the AAO, and the participating institutions. The GAMA website is \url{http://www.gama-survey.org/}.

This study made extensive use of the following tools, NumPy \citep{Harris2020}, the Tool for OPerations on Catalogues And Tables, TOPCAT \citep{Taylor2005}, and Python Data Analysis Library pandas \citep{McKinney2010}.
We would like to thank Editage (www.editage.com) for English language editing.
\end{ack}

\appendix 
\section{Model selection}\label{A1}
The proposed model has various hyperparameters such as the dimensionality of the latent space $d$, standard deviation of Gaussian noise, $r_{\rm gauss}$, and number of epochs. 
One of the challenges in machine learning methods is to select the best hyperparameters, i.e., to select the best model. In this study, we considered the hyperparameters given in table~\ref{tab;hyperparameters}.

\begin{table}
  \tbl{Hyperparameters of the proposed model}{%
  \begin{tabular}{cr}
      \hline
      parameter & tried values\\ 
      \hline
      Latent space dimension $d$ & $4, 8, 16, 32$ \\
      GaussianNoise rate $r_{\rm gauss}$ & $0, 0.01, 0.02, 0.03, 0.05$  \\
      Epoch number\footnotemark[$*$] & $10,11,\dots,25$   \\
      Iteration\footnotemark[$\dag$] & $1,2,\dots,30$    \\
      \hline
    \end{tabular}}
\label{tab;hyperparameters}
\begin{tabnote}
\footnotemark[$*$] Epoch numbers $1$--$9$ were excluded because they were at an insufficient stage of learning.\\ 
\footnotemark[$\dag$] Under the same condition, $d=8, r_{\rm gauss}=0.02$. 
\end{tabnote}
\end{table}

We notice that quantitatively varied results are obtained in different runs because the machine learning model is stochastic. 
However, the general trends as discussed in section~\ref{s4} remain unchanged. 
In addition, we noticed that variations in the results of multiple runs under a fixed set of hyperparameter values were larger than the systematic variations caused by different hyperparameter sets. 
Therefore, the quantitative evaluation of the hyperparameters is challenging. 

\begin{figure*}
 \begin{center}
  \includegraphics[width=14cm]{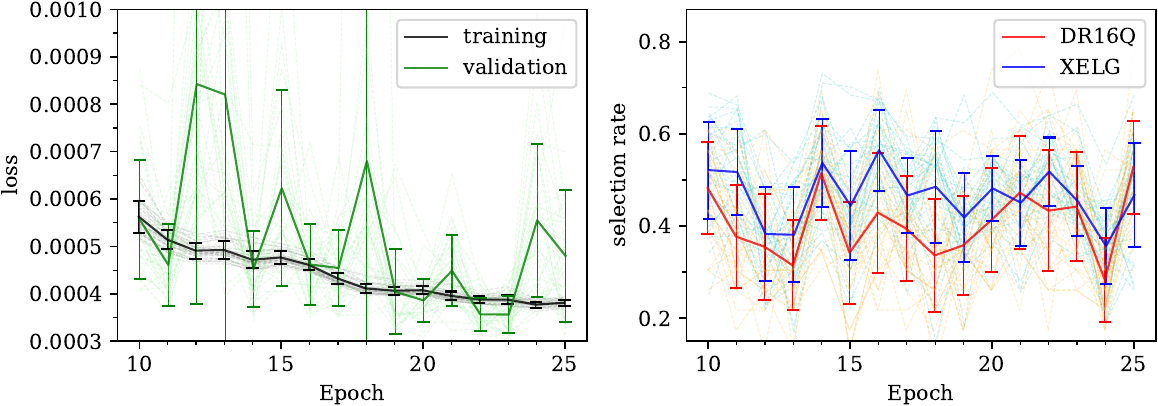} 
 \end{center}
\caption{(Left) Loss values in 480 (16 epochs in 10--25th epochs $\times$ 30 runs) various weight sets.
The black and green colors correspond to the losses for training and validation data (the original training data-set in table~\ref{data_summary} is divided into training and validation data in the ratio 4:1).
The light-colored dashed lines indicate the results of 30 runs for each epoch, and the dark-colored solid lines and error-bars depict the mean values and standard deviations in 30 runs. 
(Right) Selection rates for the DR16Q and XELG samples in 480 various weight sets.
The red and blue colors indicate the selection rates for the DR16Q and XELG, respectively.
The light-colored dashed lines indicate the results of 30 runs for each epoch, and the dark-colored solid lines and error-bars depict the mean values and standard deviations in 30 runs.
}
\label{sel_rate}
\end{figure*}

Hence, in this study, we decided to train the model several times with fixed $d=8$ and $r_{\rm gauss}=0.02$, and subsequently selected the most suitable model among them. 
These fixed parameters were selected because they provide a relatively lower average loss after ten runs as compared to the additional parameter values in table~\ref{tab;hyperparameters}. 
The model training was executed 30 times under the same condition, and the weight data from epochs 10 to 25 was saved. Thereafter, we selected the most suitable weight data from 480 (16 epochs $\times$ 30 runs) various weight data.
Loss values and selection rates calculated with 480 weight parameter sets are summarized in figure~\ref{sel_rate}.

In the model selection, we defined the most suitable weight data as the data for which the sum of the selection rates of known DR16Q quasars and XELGs, $r_{\rm sel, DR16Q} + r_{\rm sel, XELG}$ (subsection~\ref{s22} for the sample definition) was the highest.
All the results presented in the main text, including figures~\ref{activation_map} and \ref{residuals}, are based on the model with the most suitable weights. 

\section{Objects with negative residuals}\label{A2}

We examined the objects with a large $S_{\rm anom}$ and a negative residual. 
These objects were not selected as anomaly candidates and were determined more frequently in the $r$-band. 
A large $S_{\rm anom}$ with a negative residual in a certain band means a nuclear depression in that band as compared to typical values inferred from the training sample, and therefore, provides significant negative residuals in the nuclear region. 

\begin{figure*}
 \begin{center}
  \includegraphics[width=17cm]{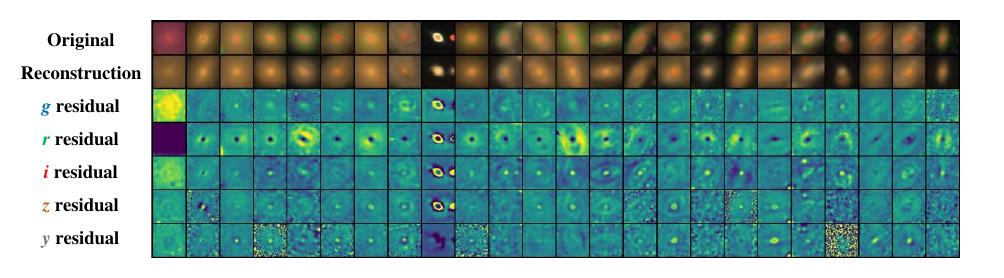} 
 \end{center}
\caption{
Original images (top row), reconstructed images (2nd row), and residual maps in the $g,r,i,z,y$ bands (3rd--7th rows) of 25 galaxies with the largest negative residuals in the $r$ band.
}
\label{residuals_minus}
\end{figure*}

Figure~\ref{residuals_minus} depicts the original, reconstruction, and residual images of sources with a negative residual and the largest $S_{\rm anom}$ in the $r$ band.
Almost all the sources were determined to have a positive residual in the central region in all the bands except the $r$ band. 
We also observe that flux counts in the $r$ band, including background levels, tend to be higher compared to those in the general field, which produce unrealistic green halos as seen in the color images (figure~\ref{residuals_minus}). 
These results indicate that a large $S_{\rm anom}$ with a negative residual in the $r$ band is likely due to reduction or technical errors rather than scientific factors, for instance, an error of zero-point correction in the $r$ band or saturation in the $i$ band in flux normalization.
Such artifact-like objects are found in the anomaly candidates as contaminants (subsection~\ref{s44}). 

\section{Further application}\label{A3}

In addition, we applied the proposed method to color-selected objects in the local universe without a spectroscopic redshift measurement conducted as a supplemental research. 
We perform a simple $grz$ color selection ($g-r<1.2$ and $r-z<1$; see also $BRI$ color selectin in the DEEP2 galaxy redshift survey, \cite{Newman2013}) to remove background sources at $z\ge0.2$ as many as possible while keeping a high selection completeness at the target redshift range of $z=0.05$--0.2. 
Although more aggressive color criteria or photo-$z$ selection allows a more strict redshift cut, it is not ideal for the deep anomaly detection: as we search for galaxies with anomalous SEDs, such high restrictions may remove potential anomaly candidates in the preselection. 

\begin{figure}
 \begin{center}
  \includegraphics[width=7.5cm]{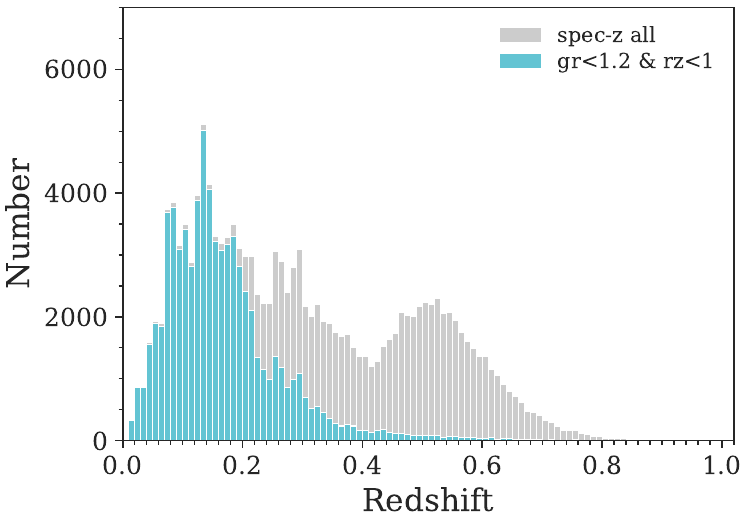} 
 \end{center}
\caption{
Spectroscopic redshift distribution of the original, the $i$ band limited sample 
(grey histogram) and the color-selected subsample (blue).
}
\label{zdist_grz}
\end{figure}

Based on the $grz$ color selection, the 839~908 of the original samples with $i<20$ mag was reduced to 487~152 sources (figure~\ref{zdist_grz}).
Given the spec-$z$ sample from 
SDSS DR15 \citep{Blanton2017,Aguado2019,Bolton2012} that consists of the main sample ($r_\mathrm{petro}<17.8$; \cite{Strauss2002}) and luminous red galaxies ($r_\mathrm{petro}<19.2$; \cite{Eisenstein2001}), a sampling completeness is 96.5\%, a removal rate of $z\le0.05$ or $z\ge0.2$ sources is 88.6\%, and an outlier fraction in the selected sample is estimated to be 20.9\%, respectively. 

\begin{table*}
\caption{
A list of identification number ({\tt object\_id}), coordinates, selection bands, and anomaly scores of the over 3$\sigma$ anomaly from the additional color-selected data. 
See supplementary data 2 for the full version of this catalog (online material).
}
  \begin{tabular}{rrrrr}
      \hline
      ID & R.A. [deg] & Dec. [deg] & Band & $S_{\rm anom}^{(g)}, S_{\rm anom}^{(r)}, S_{\rm anom}^{(i)}$\\ 
      \hline
      36411444245320355 &  30.59056 & -6.67457 & $g$\&$r$\&$i$ & (0.05678, 0.05687, 0.37663)\\
      36411452835251719 &  30.57699 & -6.38626 & $i$ & (0.00299, 0.00687, 0.02174)\\
      36411577389302008 &  30.34561 & -6.90563 & $g$ & (0.16888, 0.00131, 0.00347)\\
      36411590274208996 &  30.47591 & -6.27652 & $r$ & (0.00419, 0.08369, 0.00053)\\
      36411594569174631 &  30.52445 & -6.16192 & $i$ & (0.00106, 0.00612, 0.02948)\\
      ... & ... & ... & ... & ...\\
      \hline
    \end{tabular}
\label{tab;catalog2}
\end{table*}

\begin{figure*}
 \begin{center}
  \includegraphics[width=17cm]{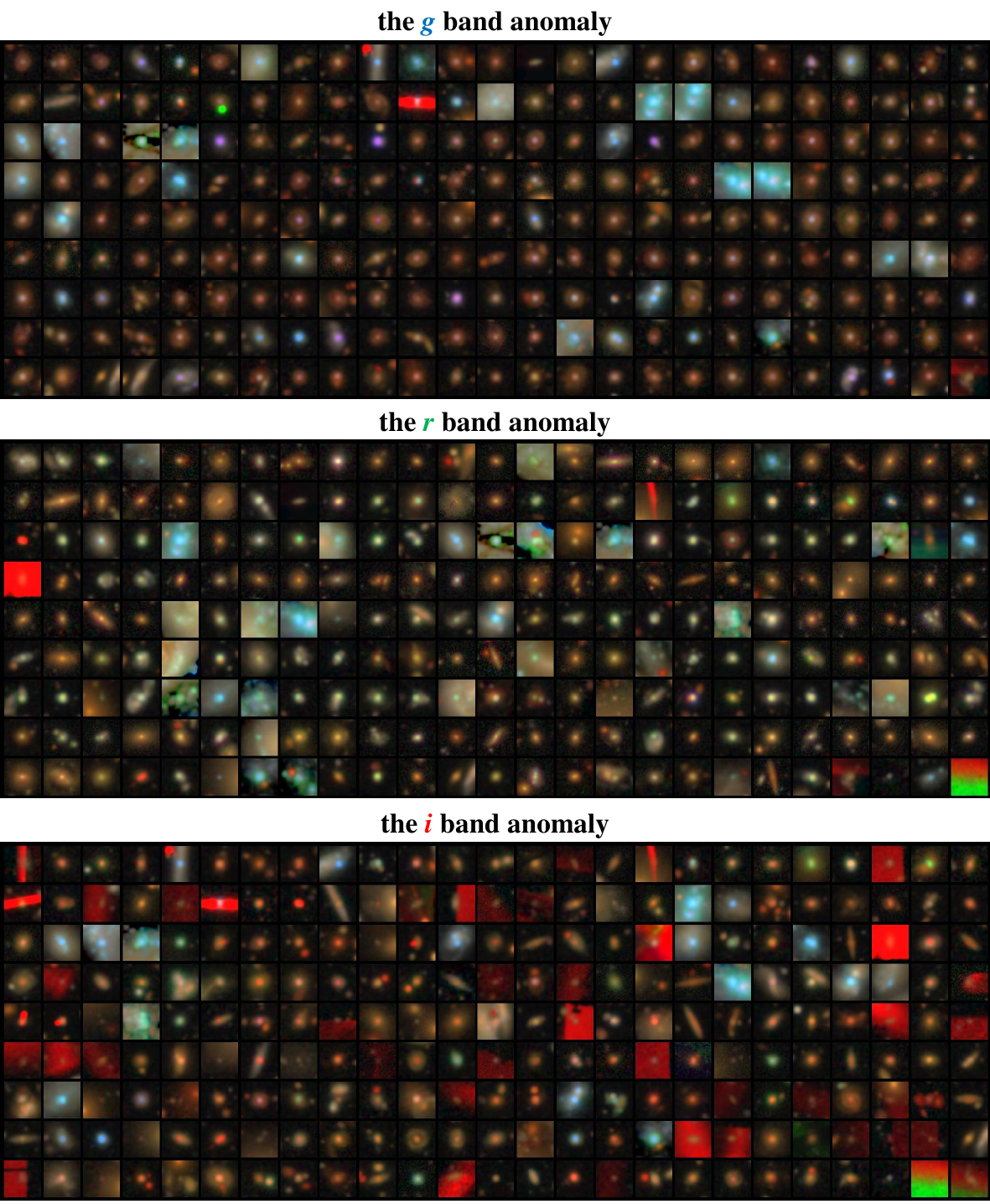} 
 \end{center}
\caption{
Postage stamps of anomaly candidates with the top 
$0.0465\%$ $S_{\rm anom}$ (225 objects per band; correspond to $3.5\sigma$ upper limit) from the additional color-selected samples in the $g,r,$ and $i$ bands.
}
\label{anomaly_additional}
\end{figure*}

Figure~\ref{anomaly_additional} represents the most anomalous sources with the top $0.0465\%$ $S_{\rm anom}$ in the color-selected objects, and their identification numbers ({\tt object\_id}), sky coordinates in the HSC-SSP PDR3 \citep{Aihara2021}, and anomaly scores are summarised in table \ref{tab;catalog2}.
In figure~\ref{anomaly_additional}, there are many red compact sources unlike our main anomaly samples (figure~\ref{anomaly_3sigma}).
They could be higher-$z$ objects because our machine-learning model has not well learned $z>0.2$ galaxies. 
On the other hand, we detect highly blue, green, or purple objects, which would be new candidates of XELGs without spectroscopic confirmations. 
There are also some clear artifacts as discussed in appendix~\ref{A2}.
We concluded that the proposed methods are also useful for detecting outliers from the color-selected sample, though further improvements of both training data-set and machine learning model are needed toward practical use.

\begin{thebibliography}{}
\bibitem[Abadi et al.(2016)]{Abadi2016}
  Abadi, M., et al.\ 2016, arXiv e-prints, arXiv:1603.04467
\bibitem[Abazajian et al.(2004)]{Abazajian2004}
  Abazajian K., et al.\ 2004, AJ, 128, 502
\bibitem[Abazajian et al.(2009)]{Abazajian2009}
  Abazajian, K.~N., et al.\ 2009, \apjs, 182, 543
\bibitem[Aguado et al.(2019)]{Aguado2019} 
  Aguado D.~S., et al.\ 2019, ApJS, 240, 23
\bibitem[Ahumada et al.(2020)]{Ahumada2020} 
  Ahumada R., et al.\ 2020, \apjs, 249, 3
\bibitem[Aihara et al.(2018)]{Aihara2018} 
  Aihara, H., et al.\ 2018, \pasj, 70, S4
\bibitem[Aihara et al.(2021)]{Aihara2021} 
  Aihara, H., et al.\ 2021, arXiv e-prints, arXiv:2108.13045
\bibitem[Akeson et al.(2019)]{Akeson2019} 
  Akeson, R., et al.\ 2019, arXiv e-prints, arXiv:1902.05569
\bibitem[Akiyama et al.(2015)]{Akiyama2015} 
  Akiyama, M., et al.\ 2015, \pasj, 67, 82
\bibitem[Amendola et al.(2018)]{Amendola2018} 
  Amendola, L., et al.\ 2018, Living Reviews in Relativity, 21, 2
\bibitem[Amor\'{i}n et al.(2015)]{Amorin2015} 
  Amor\'{i}n, R., et al.\ 2015, \aap, 578, A105
\bibitem[Andrae et al.(2010)]{Andrae2010} 
  Andrae, R., Melchior, P., \& Bartelmann, M.\ 2010, \aap, 522, A21
\bibitem[Appenzeller et al.(1998)]{Appenzeller1998} 
  Appenzeller I., et al.\ 1998, \apjs, 117, 319
\bibitem[Baldry et al.(2010)]{Baldry2010} 
  Baldry, I. K., et al.\ 2010, \mnras, 404, 86
\bibitem[Baldry et al.(2014)]{Baldry2014} 
  Baldry I.~K., et al.\ 2014, MNRAS, 441, 2440
\bibitem[Baldry et al.(2018)]{Baldry2018} 
  Baldry I.~K., et al.\ 2018, MNRAS, 474, 3875
\bibitem[Baldwin et al.(1981)]{Baldwin1981} 
  Baldwin, J.~A., Phillips, M.~M., \& Terlevich, R.\ 1981, \pasp, 93, 5
\bibitem[Baron and Poznanski(2017)]{Baron2017} 
  Baron, D., \& Poznanski, D.\ 2017, \mnras, 465, 4530
\bibitem[Bauer et al.(2000)]{Bauer2000} 
  Bauer F.E., et al.\ 2000, \apjs, 129, 547
\bibitem[Baur et al.\ (2021)]{Baur2021} 
  Baur C., et al. (2021). Medical Image Analysis, 69
\bibitem[Bisong(2019)]{Bisong2019} 
  Bisong, E.\ 2019, Google Colaboratory. in Building Machine Learning and Deep Learning Models on Google Cloud Platform (ed. Bisong, E.) 59-64
\bibitem[Bland-Hawthorn and Gerhard(2016)]{Bland2016} 
  Bland-Hawthorn, J., \& Gerhard, O.\ 2016, \araa, 54, 529
\bibitem[Blanton et al.(2017)]{Blanton2017} 
  Blanton M.~R., et al.\ 2017, AJ, 154, 28
\bibitem[Blecha et al.(2018)]{Blecha2018} 
  Blecha, L., Snyder, G. F., Satyapal, S., \& Ellison, S. L.,\ 2018, \mnras, 478, 3056
\bibitem[Bolton et al.(2012)]{Bolton2012} 
  Bolton, A.~S., et al.\ 2012, \aj, 144, 144
\bibitem[Bosch et al.(2018)]{Bosch2018} 
  Bosch J., et al.\ 2018, PASJ, 70, S5
\bibitem[Brinchmann et al.(2004)]{Brinchmann2004} 
  Brinchmann, J., et al.\ 2004, \mnras, 351, 1151
\bibitem[Bundy et al.(2015)]{Bundy2015}
  Bundy, K., et al.\ 2015, \apj, 798, 7
\bibitem[Cardamone et al.(2009)]{Cardamone2009} 
  Cardamone, C., et al.\ 2009, \mnras, 399, 1191
\bibitem[Chang et al.(2019)]{Chang2019} 
  Chang, Y. -L., et al.\ 2019, \aap, 632A, 77
\bibitem[Chen et al.(2018)]{Chen2018} 
  Chen, X., \& Konukoglu, E.\ 2018,  arXiv e-prints, arXiv:1806.04972.
\bibitem[Chollet et al.(2015)]{Chollet2015} 
  Chollet, F., et al.\ 2015, Available at: \url{https://keras.io}
\bibitem[Colless et al.(2001)]{Colless2001} 
  Colless M., et al.\ 2001, \mnras, 328, 1039
\bibitem[Coupon et al.(2018)]{Coupon2018} 
  Coupon J., et al.\ 2018, PASJ, 70, S7
\bibitem[Croom et al.(2004)]{Croom2004} 
  Croom S.M., et al.\ 2004, \mnras, 349, 1397
\bibitem[Croom et al.(2009)]{Croom2009} 
  Croom, S. M. et al.\ 2009, \mnras, 392, 19
\bibitem[Dong et al.(2018)]{Dong2018} 
  Dong X.Y., et al.\ 2018, \aj, 155, 189
\bibitem[Eisenstein et al.(2001)]{Eisenstein2001}
  Eisenstein, D.~J., et al.\ 2001, \aj, 122, 2267
\bibitem[Esquej et al.(2013)]{Esquej2013} 
  Esquej, P., et al.\ 2013, \aap, 557, 123
\bibitem[Flesch(2021)]{Flesch2021} 
  Flesch, E.~W\ 2021, arXiv e-prints, arXiv:2105.12985
\bibitem[Flesch(2019)]{Flesch2019} 
  Flesch, E.~W\ 2021, arXiv e-prints, arXiv:1912.05614
\bibitem[Foltz et al.(1989)]{Foltz1989} 
  Foltz C.B., et al.\ 1989, \aj, 98, 1959
\bibitem[Francis et al.(2004)]{Francis2004} 
  Francis P.J., Nelson B.O., \& Cutri R.M.\ 2004, \aj, 127, 646
\bibitem[Furusawa et al.(2018)]{Furusawa2018} 
  Furusawa, H., et al.\ 2018, \pasj, 70, S3
\bibitem[Fustes et al.(2013)]{Fustes2013} 
  Fustes, D., et al.\ 2013, \aap, 559, A7
\bibitem[Garilli et al.(2014)]{Garilli2014} 
  Garilli, B., et al.\ 2014, \aap, 562, 23
\bibitem[Ge et al.(2012)]{Ge2012} 
  Ge J.-Q., et al.\ 2012, \apjs, 201, 31
\bibitem[Glorot et al.(2011)]{Glorot2011} 
  Glorot, X., Bordes, A., \& Bengio. Y.\ 2011, In Proc. 14th International Conf. on Artificial Intelligence and Statistics, 315, 323
\bibitem[Gosset et al.(1997)]{Gosset1997} 
  Gosset, E., et al.\ 1997, \aaps, 123, 529
\bibitem[Greis at al.(2016)]{Greis2016} 
  Greis, S. M. L., et al.\ 2016, \mnras, 459, 2591
\bibitem[Harikane et al.(2019)]{Harikane2019} 
  Harikane Y., et al.,\ 2019, \apj, 883, 142
\bibitem[Harris et al.(2020)]{Harris2020} 
  Harris, C. R., et al.\ 2020, Nature, 585, 357
\bibitem[He and Chen(1993)]{He1993} 
  He R., \& Chen J.-S.\ 1993, \apss, 200, 279
\bibitem[Hinton and Salakhutdinov(2006)]{Hinton2006} 
  Hinton, G.~E., \& Salakhutdinov, R.~R.\ 2006, Science, 313, 504
\bibitem[Hocking et al.(2018)]{Hocking2018} 
  Hocking, A., et al.\ 2018, \mnras, 473, 1108]
\bibitem[Hopkins et al.(2008)]{Hopkins2008} 
  Hopkins, P. F., Cox, T. J., Kere\v{s}, D., \& Hernquist, L.\ 2008, \apjs, 175, 390
\bibitem[Hopkins et al.(2013)]{Hopkins2013} 
  Hopkins A.~M., et al.\ 2013, MNRAS, 430, 2047
\bibitem[Ioffe and Szegedy(2015)]{Ioffe2015} 
  Ioffe, S., \& Szegedy, C.\ 2015, arXiv e-prints, arXiv:1502.03167
\bibitem[Ishino et al.(2020)]{Ishino2020} 
  Ishino, T., et al.\ 2020, \pasj, 72, 83
\bibitem[Ivezi\'{c} et al.(2019)]{Ivezic2019} 
  Ivezi\'{c}, Z., et al.\ 2019, \apj , 873, 111
\bibitem[Kauffmann et al.(2003)]{Kauffmann2003} 
  Kauffmann, G., et al.\ 2003, \mnras, 341, 33
\bibitem[Kawanomoto et al.(2018)]{Kawanomoto2018} 
  Kawanomoto, S., et al.\ 2018, \pasj, 70, 66
\bibitem[Kingma and Ba(2014)]{Kingma2014} 
  Kingma, D.~P., \& Ba, J.\ 2014, arXiv e-prints, arXiv:1412.6980
\bibitem[Kojima et al.(2020)]{Kojima2020} 
  Kojima, T., et al.\ 2020, \apj, 898, 142
\bibitem[Komatsu et al.(2011)]{Komatsu2011} 
  Komatsu, E., et al.\ 2011, \apjs, 192, 18
\bibitem[Komiyama et al.(2018)]{Komiyama2018} 
  Komiyama, Y., et al.\ 2018, \pasj, 70, S2
\bibitem[Kormendy and Ho(2013)]{Kormendy2013} 
  Kormendy, J., \& Ho, L.~C.\ 2013, \araa, 51, 511
\bibitem[Krizhevsky et al.(2012)]{Krizhevsky2012} 
  Krizhevsky, A., Sutskever, I., \& Hinton, G.\ 2012, In Proc. Advances in Neural Information Processing Systems, 25, 1090-1098
\bibitem[Kroupa(2001)]{Kroupa2001}
  Kroupa, P.\ 2001, \mnras, 322, 231
\bibitem[Lacy et al(2007)]{Lacy2007} 
  Lacy, M., et al.\ 2007, \aj, 133, 186
\bibitem[Lacy et al.(2013)]{Lacy2013} 
  Lacy, M., et al.\ 2013, \apjs, 208, 24
\bibitem[Lecun et al.(1998)]{Lecun1998} 
  Lecun Y., et al.\ 1998, Proc. IEEE, 86, 2278
\bibitem[Lecun et al.(2015)]{Lecun2015} 
  Lecun, Y., Bengio, Y., \& Hinton, G.\ 2015, \nat, 521, 436
\bibitem[Li et al(2021)]{Li2021} 
  Li, J., et al.\ 2021, arXiv e-prints, arXiv:2105.06568
https://arxiv.org/abs/2105.06568
\bibitem[Lintott et al.(2008)]{Lintott2008} 
  Lintott C.~J., et al.\ 2008, MNRAS, 389, 1179
\bibitem[Lintott et al.(2011)]{Lintott2011} 
  Lintott C., et al.\ 2011, MNRAS, 410, 166
\bibitem[Liu et al.(2019)]{Liu2019} 
  Liu H.-Y., et al.\ 2019, \apjs, 243, 21
\bibitem[Lupton et al.(1999)]{Lupton1999} 
  Lupton, R.~H., Gunn, J.~E., \& Szalay, A.~S.\ 1999, \aj, 118, 1406
\bibitem[Lupton et al.(2004)]{Lupton2004} 
  Lupton R., et al.\ 2004, \pasp, 116, 133
\bibitem[Lyke et al.(2020)]{Lyke2020} 
  Lyke, B.~W., et al.\ 2020, \apjs, 250, 8
\bibitem[MacAlpine et al.(1981)]{MacAlpine1981} 
  MacAlpine G.M., \& Williams G.A.\ 1981, \apjs, 45, 113
\bibitem[Margalef-Bentabol et al.(2020)]{Margalef2020} 
  Margalef-Bentabol B., et al.\ 2020, \mnras, 496, 2346
\bibitem[Markelijn et al.(1970)]{Markelijn1970} 
  Merkelijn J., \& Wall J.V.\ 1970, AuJPh, 23, 575
\bibitem[Massaro et al.(2009)]{Massaro2009} 
  Massaro E., et al.\ 2009, \aap, 495, 691
\bibitem[Matsuoka et al.(2014)]{Matsuoka2014} 
  Matsuoka, Y., Strauss, M. A., Price, T. N., III, \& DiDonato,M. S.\ 2014, \apj, 780, 162
\bibitem[Matsuoka et al.(2019)]{Matsuoka2019} 
  Matsuoka, Y., et al.\ 2019, \apjl, 872, L2
\bibitem[McKinney et al.(2010)]{McKinney2010} 
  McKinney W., et al.\ 2010, Proc. 9th Python Sci. Conf., 1697900, 51
\bibitem[Meusinger et al.(2012)]{Meusinger2012} 
  Meusinger, H., et al.\ 2012, \aap, 541, A77
\bibitem[Miyazaki et al.(2018)]{Miyazaki2018} 
  Miyazaki, S., et al.\ 2018, \pasj, 70, S1
\bibitem[Mortlock et al.(2012)]{Mortlock2012} 
  Mortlock, D. J., et al.\ 2012, \mnras, 419, 390
\bibitem[Naab and Ostriker(2017)]{Naab2017} 
  Naab, T., \& Ostriker, J.~P.\ 2017, \araa, 55, 59
\bibitem[Nachman(2020)]{Nachman2020} 
  Nachman, B.\ 2020, arXiv e-prints, arXiv:2010.14554
\bibitem[Newman et al.(2013)]{Newman2013} 
  Newman, J.~A., et al.\ 2013, \apjs, 208, 5
\bibitem[Oke and Gunn(1983)]{Oke1983} 
  Oke, J.~B., \& Gunn, J.~E.\ 1983, \apj, 266, 713
\bibitem[Paturel et al.(2003)]{Paturel2003} 
  Paturel, G., et al.\ 2003, \aap, 412, 45
\bibitem[Renzini and Peng(2015)]{Renzini2015}
  Renzini, A., \& Peng, Y.-j.\ 2015, \apj, 801, L29
\bibitem[Rowan-Robinson et al.(2013)]{Rowan2013} 
  Rowan-Robinson, M., et al.\ 2013, \mnras, 428, 1958
\bibitem[Rumelhart, Hinton, \& Williams(1986)]{Rumelhart1986}
  Rumelhart, D. E., Hinton, G. E., Williams, R. J.\ 1986, \nat, 323, 533
\bibitem[Salim et al.(2007)]{Salim2007} 
  Salim, S., et al.\ 2007, \apjs, 173, 267
\bibitem[Salim et al.(2016)]{Salim2016} 
  Salim S., et al., 2016, ApJS, 227, 2
\bibitem[Salim, Boquien, \& Lee(2018)]{Salim2018} 
  Salim S., Boquien M., Lee J. C., 2018, ApJ, 859, 11
\bibitem[S\'{a}nchez Almeida and Allende Prieto(2013)]{Sanchez2013} 
  S\'{a}nchez Almeida J., \& Allende Prieto C.\ 2013, \apj, 763, 50
\bibitem[Schawinski et al.(2010)]{Schawinski2010} 
  Schawinski, K., et al.\ 2010, \apj, 711, 284
\bibitem[Secrest et al.(2015)]{Secrest2015} 
  Secrest, N., et al.\ 2015, \apjs, 221, 12
\bibitem[Silverman et al.(2020)]{Silverman2020} 
  Silverman, J. D., et al.\ 2020, \apj, 899, 154
\bibitem[Solarz et al.(2017)]{Solarz2017} 
  Solarz, A., et al.\ 2017, \aap, 606, A39
\bibitem[Sonnenfeld et al.(2018)]{Sonnenfeld2018} 
  Sonnenfeld A., et al., 2018, \pasj, 70, S29
\bibitem[Stalin et al.(2010)]{Stalin2010} 
  Stalin, C. S., et al.\ 2010, \mnras, 401, 294
\bibitem[Stark et al.(2018)]{Stark2018} 
  Stark, D., et al.\ 2018, \mnras, 477, 2513
\bibitem[Storey-Fisher et al.(2021)]{Storey-Fisher2021} 
  Storey-Fisher, K., et al.\ 2021, arXiv e-prints, arXiv:2105.02434
\bibitem[Strauss et al.(2002)]{Strauss2002} 
  Strauss, M.~A., et al.\ 2002, \aj, 124, 1810
\bibitem[Takada et al.(2014)]{Takada2014} 
  Takada, M., et al.\ 2014, \pasj, 66, R1
\bibitem[Tamura et al.(2016)]{Tamura2016} 
  Tamura, N., et al.\ 2016, \procspie, 9908, 99081M
\bibitem[Tang et al(2021)]{Tang2021} 
  Tang, S., et al.\ 2021, arXiv e-prints, arXiv:2105.10163
\bibitem[Taylor(2005)]{Taylor2005} 
  Taylor, M. B.\ 2005, Astronomical Data Analysis Software and Systems XIV, 347, 29
\bibitem[Toba et al.(2015)]{Toba2015} 
 Toba, Y., et al.\ 2015, \pasj, 67, 86
\bibitem[Toba et al.(2018)]{Toba2018} 
 Toba, Y., et al.\ 2018, \apj, 857, 31
\bibitem[Toba et al(2021)]{Toba2021} 
 Toba, Y., et al.\ 2021, arXiv e-prints, arXiv:2106.14527
\bibitem[Tremonti et al.(2004)]{Tremonti2004} 
  Tremonti, C.~A., et al.\ 2004, \apj, 613, 898
\bibitem[Trichas et al.(2012)]{Trichas2012} 
  Trichas, M., et al.\ 2012, \apjs, 200, 17 
\bibitem[Trump et al.(2007)]{Trump2007} 
  Trump, J. R., et al.\ 2007, \apjs, 172, 383
\bibitem[Trump et al.(2015)]{Trump2015} 
  Trump, J. R., et al.\ 2015, \apj, 811, 26
\bibitem[Vincent et al.(2008)]{Vincent2008} 
  Vincent P., et al.\ 2008, in Proc. of ICML. 1096-1103.
\bibitem[Wechsler and Tinker(2018)]{Wechsler2018} 
  Wechsler, R.~H., \& Tinker, J.~L.\ 2018, \araa, 56, 435
\bibitem[Willett et al.(2013)]{Willett2013} 
  Willett K.~W., et al.\ 2013, \mnras, 435, 2835
\bibitem[Yang et al.(2017)]{Yang2017} 
  Yang, H., et al.\ 2017, \apj, 847, 38
\bibitem[York et al.(2000)]{York2000} 
  York, D. G., et al.\ 2000, \aj, 120, 1579
\bibitem[Zaccarelli, Bindi, and Strollo(2021)]{Zaccarelli2021} 
Zaccarelli, R., Bindi, D., \& Strollo, A.\ 2021, Seismological Research Letters, 92, 2627
\bibitem[Zhou et al.(2006)]{Zhou2006} 
  Zhou H., et al.\ 2006, \apjs, 166, 128 
\bibitem[Zimmerer et al.(2018)]{Zimmerer2018} 
  Zimmerer, D., et al.\ 2018,  arXiv e-prints, arXiv:1812.05941
\end{thebibliography}
\end{document}